\documentclass[12pt,english]{article}
\usepackage{mathptmx}
\usepackage[T1]{fontenc}
\usepackage[latin9]{inputenc}
\usepackage{geometry}
\usepackage{color}
\usepackage{babel}
\usepackage{mathtools}
\usepackage{amsmath}
\usepackage{amssymb}
\usepackage{stmaryrd}
\usepackage{graphicx}
\usepackage{setspace}
\usepackage[authoryear]{natbib}
\usepackage[unicode=true,
 bookmarks=true,bookmarksnumbered=false,bookmarksopen=false,
 breaklinks=false,pdfborder={0 0 1},backref=false,colorlinks=true]
 {hyperref}
\hypersetup{
 citecolor=blue,linkcolor=blue}
\usepackage{breakurl}

\makeatletter

\providecommand{\tabularnewline}{\\}

\@ifundefined{date}{}{\date{}}
\usepackage{babel}

\usepackage{times}
\usepackage{pdfsync}
\usepackage[font={small}]{caption}

\bibpunct{(}{)}{,}{a}{,}{,}

\makeatother

\begin{document}
\global\long\def\real{\mathbb{R}}
 \global\long\def\RefVol{\Omega_{0}}
 \global\long\def\Refx{\mathbf{X}}
 \global\long\def\Curx{\mathbf{x}}
 \global\long\def\map{\boldsymbol{\chi}}
 \global\long\def\defgrad{\mathbf{F}}
 \global\long\def\defgradT{\mathbf{F}^{\mathrm{T}}}
 \global\long\def\defgradTi{\mathbf{F}^{\mathrm{-T}}}
 \global\long\def\d{\mathrm{d}}
 \global\long\def\RCG{\mathbf{C}}
 \global\long\def\LCG{\mathbf{b}}
 \global\long\def\div#1{\nabla\cdot#1}
 \global\long\def\curl#1{\nabla\times#1}
 \global\long\def\T#1{#1^{\mathrm{T}}}
 \global\long\def\CurStress{\boldsymbol{\sigma}}
 \global\long\def\Refcurl#1{\nabla_{\mathbf{X}}\times#1}
 \global\long\def\s#1{#1^{\star}}
 \global\long\def\m#1{#1^{(m)}}
 \global\long\def\f#1{#1^{(f)}}
 \global\long\def\p#1{#1^{(p)}}
 \global\long\def\Refdiv#1{\nabla_{\mathbf{X}}\cdot#1}
 \global\long\def\phase#1{#1^{\left(p\right)}}
 \global\long\def\jm{j_{m}^{(p)}}
 \global\long\def\xinc{\mathbf{\dot{x}}}
 \global\long\def\Pinc{\mathbf{\dot{P}}}
 \global\long\def\Einc{\mathbf{\dot{E}}}
 \global\long\def\Finc{\mathbf{\dot{F}}}
 \global\long\def\dinc{\mathbf{\check{d}}}
 \global\long\def\einc{\mathbf{\check{e}}}
 \global\long\def\Dinc{\mathbf{\dot{D}}}
 \global\long\def\sinc{\boldsymbol{\Sigma}}
 \global\long\def\fA{\mathbb{\boldsymbol{\mathscr{A}}}}
 \global\long\def\fB{\mathbb{\boldsymbol{\mathscr{B}}}}
 \global\long\def\fC{\mathbb{\boldsymbol{\mathscr{C}}}}
 \global\long\def\plane{\left(x_{1},x_{3}\right)}
 \global\long\def\anti{\dot{x}_{2}}
 \global\long\def\G{\mathbf{G}}
 \global\long\def\Gt{\mathbf{G}'}
 \global\long\def\k{\mathbf{k}}
 \global\long\def\Gtk{\left(\Gt+\k\right)}
 \global\long\def\Gk{\left(\G+\k\right)}
 \global\long\def\GGtk{\left(\G+\Gt+\k\right)}
 \global\long\def\ebias{\hat{e}}
 \global\long\def\deformation{\boldsymbol{\chi}}
 \global\long\def\dg{\mathbf{F}}
 \global\long\def\dgcomp#1{F_{#1}}
 \global\long\def\piola{\mathbf{P}}
 \global\long\def\refbody{\Omega_{0}}
 \global\long\def\refbnd{\partial\refbody}
 \global\long\def\bnd{\partial\Omega}
 \global\long\def\rcg{\mathbf{C}}
 \global\long\def\lcg{\mathbf{b}}
 \global\long\def\rcgcomp#1{C_{#1}}
 \global\long\def\cronck#1{\delta_{#1}}
 \global\long\def\lcgcomp#1{b_{#1}}
 \global\long\def\deformation{\boldsymbol{\chi}}
 \global\long\def\dgt{\dg^{\mathrm{T}}}
 \global\long\def\idgcomp#1{F_{#1}^{-1}}
 \global\long\def\velocity{\mathbf{v}}
 \global\long\def\accel{\mathbf{a}}
 \global\long\def\vg{\mathbf{l}}
 \global\long\def\idg{\dg^{-1}}
 \global\long\def\cauchycomp#1{\sigma_{#1}}
 \global\long\def\idgt{\dg^{\mathrm{-T}}}
 \global\long\def\cauchy{\boldsymbol{\sigma}}
 \global\long\def\normal{\mathbf{n}}
 \global\long\def\normall{\mathbf{N}}
 \global\long\def\traction{\mathbf{t}}
 \global\long\def\tractionl{\mathbf{t}_{L}}
 \global\long\def\ed{\mathbf{d}}
 \global\long\def\edcomp#1{d_{#1}}
 \global\long\def\edl{\mathbf{D}}
 \global\long\def\edlcomp#1{D_{#1}}
 \global\long\def\ef{\mathbf{e}}
 \global\long\def\efcomp#1{e_{#1}}
 \global\long\def\efl{\mathbf{E}}
 \global\long\def\freech{q_{e}}
 \global\long\def\surfacech{w_{e}}
 \global\long\def\outer#1{#1^{\star}}
 \global\long\def\perm{\epsilon_{0}}
 \global\long\def\matper{\epsilon}
 \global\long\def\jump#1{\llbracket#1\rrbracket}
 \global\long\def\identity{\mathbf{I}}
 \global\long\def\area{\mathrm{d}a}
 \global\long\def\areal{\mathrm{d}A}
 \global\long\def\refsys{\mathbf{X}}
 \global\long\def\Grad{\nabla_{\refsys}}
 \global\long\def\grad{\nabla}
 \global\long\def\divg{\nabla\cdot}
 \global\long\def\Div{\nabla_{\refsys}}
 \global\long\def\derivative#1#2{\frac{\partial#1}{\partial#2}}
 \global\long\def\aef{\Psi}
 \global\long\def\dltendl{\edl\otimes\edl}
 \global\long\def\tr#1{\mathrm{tr}#1}
 \global\long\def\ii#1{I_{#1}}
 \global\long\def\dh{\hat{D}}
 \global\long\def\inc#1{\dot{#1}}
 \global\long\def\sys{\mathbf{x}}
 \global\long\def\curl{\nabla}
 \global\long\def\Curl{\nabla_{\refsys}}
 \global\long\def\piolaincpush{\boldsymbol{\Sigma}}
 \global\long\def\piolaincpushcomp#1{\Sigma_{#1}}
 \global\long\def\edlincpush{\check{\mathbf{d}}}
 \global\long\def\edlincpushcomp#1{\check{d}_{#1}}
 \global\long\def\efincpush{\check{\mathbf{e}}}
 \global\long\def\efincpushcomp#1{\check{e}_{#1}}
 \global\long\def\elaspush{\boldsymbol{\mathcal{C}}}
 \global\long\def\elecpush{\boldsymbol{\mathcal{A}}}
 \global\long\def\elaselecpush{\boldsymbol{\mathcal{B}}}
 \global\long\def\disgrad{\mathbf{h}}
 \global\long\def\disgradcomp#1{h_{#1}}
 \global\long\def\trans#1{#1^{\mathrm{T}}}
 \global\long\def\phase#1{#1^{\left(p\right)}}
 \global\long\def\elecpushcomp#1{\mathcal{A}_{#1}}
 \global\long\def\elaselecpushcomp#1{\mathcal{B}_{#1}}
 \global\long\def\elaspushcomp#1{\mathcal{C}_{#1}}
 \global\long\def\dnh{\aef_{DH}}
 \global\long\def\woo{\varsigma}
 \global\long\def\wif{\Lambda}
 \global\long\def\structurefun{S}
 \global\long\def\dg{\mathbf{F}}
 \global\long\def\dgcomp#1{F_{#1}}
 \global\long\def\piola{\mathbf{P}}
 \global\long\def\refbody{\Omega_{0}}
 \global\long\def\refbnd{\partial\refbody}
 \global\long\def\bnd{\partial\Omega}
 \global\long\def\rcg{\mathbf{C}}
 \global\long\def\lcg{\mathbf{b}}
 \global\long\def\rcgcomp#1{C_{#1}}
 \global\long\def\cronck#1{\delta_{#1}}
 \global\long\def\lcgcomp#1{b_{#1}}
 \global\long\def\deformation{\boldsymbol{\chi}}
 \global\long\def\dgt{\dg^{\mathrm{T}}}
 \global\long\def\idgcomp#1{F_{#1}^{-1}}
 \global\long\def\velocity{\mathbf{v}}
 \global\long\def\accel{\mathbf{a}}
 \global\long\def\vg{\mathbf{l}}
 \global\long\def\idg{\dg^{-1}}
 \global\long\def\cauchycomp#1{\sigma_{#1}}
 \global\long\def\idgt{\dg^{\mathrm{-T}}}
 \global\long\def\cauchy{\boldsymbol{\sigma}}
 \global\long\def\normal{\mathbf{n}}
 \global\long\def\normall{\mathbf{N}}
 \global\long\def\traction{\mathbf{t}}
 \global\long\def\tractionl{\mathbf{t}_{L}}
 \global\long\def\ed{\mathbf{d}}
 \global\long\def\edcomp#1{d_{#1}}
 \global\long\def\edl{\mathbf{D}}
 \global\long\def\edlcomp#1{D_{#1}}
 \global\long\def\ef{\mathbf{e}}
 \global\long\def\efcomp#1{e_{#1}}
 \global\long\def\efl{\mathbf{E}}
 \global\long\def\freech{q_{e}}
 \global\long\def\surfacech{w_{e}}
 \global\long\def\outer#1{#1^{\star}}
 \global\long\def\perm{\epsilon_{0}}
 \global\long\def\matper{\epsilon}
 \global\long\def\jump#1{\llbracket#1\rrbracket}
 \global\long\def\identity{\mathbf{I}}
 \global\long\def\area{\mathrm{d}a}
 \global\long\def\areal{\mathrm{d}A}
 \global\long\def\refsys{\mathbf{X}}
 \global\long\def\Grad{\nabla_{\refsys}}
 \global\long\def\grad{\nabla}
 \global\long\def\divg{\nabla\cdot}
 \global\long\def\Div{\nabla_{\refsys}}
 \global\long\def\derivative#1#2{\frac{\partial#1}{\partial#2}}
 \global\long\def\aef{\Psi}
 \global\long\def\dltendl{\edl\otimes\edl}
 \global\long\def\tr#1{\mathrm{tr}\left(#1\right)}
 \global\long\def\ii#1{I_{#1}}
 \global\long\def\dh{\hat{D}}
 \global\long\def\lage{\mathbf{E}}
 \global\long\def\inc#1{\dot{#1}}
 \global\long\def\sys{\mathbf{x}}
 \global\long\def\curl{\nabla}
 \global\long\def\Curl{\nabla_{\refsys}}
 \global\long\def\piolaincpush{\boldsymbol{\Sigma}}
 \global\long\def\piolaincpushcomp#1{\Sigma_{#1}}
 \global\long\def\edlincpush{\check{\mathbf{d}}}
 \global\long\def\edlincpushcomp#1{\check{d}_{#1}}
 \global\long\def\efincpush{\check{\mathbf{e}}}
 \global\long\def\efincpushcomp#1{\check{e}_{#1}}
 \global\long\def\elaspush{\boldsymbol{\mathcal{C}}}
 \global\long\def\elecpush{\boldsymbol{\mathcal{A}}}
 \global\long\def\elaselecpush{\boldsymbol{\mathcal{B}}}
 \global\long\def\disgrad{\mathbf{h}}
 \global\long\def\disgradcomp#1{h_{#1}}
 \global\long\def\trans#1{#1^{\mathrm{T}}}
 \global\long\def\phase#1{#1^{\left(p\right)}}
 \global\long\def\elecpushcomp#1{\mathcal{A}_{#1}}
 \global\long\def\elaselecpushcomp#1{\mathcal{B}_{#1}}
 \global\long\def\elaspushcomp#1{\mathcal{C}_{#1}}
 \global\long\def\dnh{\aef_{DG}}
 \global\long\def\dnhc{\mu\lambda^{2}}
 \global\long\def\dnhcc{\frac{\mu}{\lambda^{2}}+\frac{1}{\matper}d_{2}^{2}}
 \global\long\def\dnhb{\frac{1}{\matper}d_{2}}
 \global\long\def\afreq{\omega}
 \global\long\def\dispot{\phi}
 \global\long\def\edpot{\varphi}
 \global\long\def\kh{\hat{k}}
 \global\long\def\afreqh{\hat{\afreq}}
 \global\long\def\phasespeed{c}
 \global\long\def\bulkspeed{c_{B}}
 \global\long\def\speedh{\hat{c}}
 \global\long\def\dhth{\dh_{th}}
 \global\long\def\bulkspeedl{\bulkspeed_{\lambda}}
 \global\long\def\khth{\hat{k}_{th}}
 \global\long\def\p#1{#1^{\left(p\right)}}
 \global\long\def\maxinccomp#1{\inc{\outer{\sigma}}_{#1}}
 \global\long\def\maxcomp#1{\outer{\sigma}_{#1}}
 \global\long\def\relper{\matper_{r}}
 \global\long\def\sdh{\hat{d}}
 \global\long\def\iee{\varphi}
 \global\long\def\effectivemu{\tilde{\mu}}
 \global\long\def\fb#1{#1^{\left(a\right)}}
 \global\long\def\mt#1{#1^{\left(b\right)}}
 \global\long\def\phs#1{#1^{\left(p\right)}}
 \global\long\def\thc{h}
 \global\long\def\state{\mathbf{s}}
 \global\long\def\harmonicper{\breve{\matper}}
 \global\long\def\kb{k_{B}}
 \global\long\def\cb{\bar{c}}
 \global\long\def\mb{\bar{\mu}}
 \global\long\def\rb{\bar{\rho}}
 \global\long\def\wavenumber{k}
 \global\long\def\nh{\mathbf{n}}
 \global\long\def\mh{\mathbf{m}}
 \global\long\def\deflect{\inc x_{2}}
 \global\long\def\sdd#1{#1_{2,11}}
 \global\long\def\sdddd#1{#1_{2,1111}}
 \global\long\def\sd#1{#1_{2,1}}
 \global\long\def\sddd#1{#1_{2,111}}
 \global\long\def\xdddd#1{#1_{,\xi\xi\xi\xi}}
 \global\long\def\xdd#1{#1_{,\xi\xi}}
 \global\long\def\xd#1{#1_{,\xi}}
 \global\long\def\xddd#1{#1_{,\xi\xi\xi}}
 \global\long\def\jm{J_{m}}
 \global\long\def\dv{\Delta V}
 \global\long\def\ih{\mathbf{i}_{1}}
 \global\long\def\kh{\mathbf{i}_{3}}
 \global\long\def\jh{\mathbf{i}_{2}}
 \global\long\def\etil{E}
 \global\long\def\genT{\mathsf{Q}}
 \global\long\def\transfer{\mathsf{T}}
 \global\long\def\statevec{\mathbf{s}}
 \global\long\def\coefvec{\mathbf{c}}
 \global\long\def\pressure{p_{0}}
 \global\long\def\ncell#1{#1_{\left(n\right)}}
 \global\long\def\ydisp{\inc x_{2}}
 \global\long\def\ycord{x_{2}}
 \global\long\def\pn#1{\ncell{#1}^{\left(p\right)}}
 \global\long\def\pnm#1{#1_{\left(n\right)m}^{\left(p\right)}}
 \global\long\def\eigen{\alpha}
 \global\long\def\xcomp{x_{1}}
 \global\long\def\totalT{\mathsf{T_{\mathrm{tot}}}}
 \global\long\def\rads{\frac{\mathrm{rad}}{\mathrm{s}}}
 \global\long\def\lf{\gamma}
 \global\long\def\tf{T_{m}}
 \global\long\def\eigenim{\beta}
 \global\long\def\bS{\mathsf{S}}

\global\long\def\Blochwn{k_{B}}
 \global\long\def\waven{k}

\global\long\def\Beff{\tilde{B}}
 \global\long\def\rhoeff{\tilde{\rho}}
 \global\long\def\aveM{\overline{M}}
 \global\long\def\avep{\overline{p}}
 \global\long\def\aveu{\overline{u}}
 \global\long\def\avevarsigma{\overline{\varsigma}}

\global\long\def\uF{u_{\mathrm{F}}}
 \global\long\def\wF{w_{\mathrm{F}}}
 \global\long\def\rhoF{\rho_{\mathrm{F}}}
 \global\long\def\BF{B_{\mathrm{F}}}
 \global\long\def\dl{l^{(i)}}

\global\long\def\bF{\state_{\mathrm{d}}}
 \global\long\def\bA{\state_{\mathrm{f}}}
 \global\long\def\ba{\coefvec}
 \global\long\def\bcero{\mathsf{0}}
 \global\long\def\bH{\mathsf{H}}
 \global\long\def\ltotal{l_{\mathrm{tot}}}
 \global\long\def\bI{\mathsf{I}_{2}}
 \global\long\def\qpd{\genT_{\mathrm{d}}^{+}}
 \global\long\def\qmd{\genT_{\mathrm{d}}^{-}}
 \global\long\def\qpf{\genT_{\mathrm{f}}^{+}}
 \global\long\def\qmf{\genT_{\mathrm{f}}^{-}}
 \global\long\def\curcon{\Omega}
 \global\long\def\ld{\mathbf{D}}
 \global\long\def\qpdz{\genT_{\mathrm{d}\left(0\right)}^{+}}
 \global\long\def\qmdz{\genT_{\mathrm{d}\left(0\right)}^{-}}
 \global\long\def\qpfz{\genT_{\mathrm{f}\left(0\right)}^{+}}
 \global\long\def\qmfz{\genT_{\mathrm{f}\left(0\right)}^{-}}
 \global\long\def\qpdl{\genT_{\mathrm{d}\left(M\right)}^{+}}
 \global\long\def\qmdl{\genT_{\mathrm{d}\left(M\right)}^{-}}
 \global\long\def\qpfl{\genT_{\mathrm{f}\left(M\right)}^{+}}
 \global\long\def\qmfl{\genT_{\mathrm{f}\left(M\right)}^{-}}
 \global\long\def\e{\mathop{\rm \mbox{{\Large e}}}\nolimits}
 \global\long\def\Tr{\textrm{Tr}}
 \global\long\def\Det{\textrm{Det}}
 \global\long\def\sgn{\textrm{sgn}}
 \global\long\def\pr{^{\prime}}
 \global\long\def\bn#1{\mbox{\boldmath\ensuremath{#1}}}
 \global\long\def\bB{\bn B}
 \global\long\def\bP{\bn P}
 \global\long\def\bY{\bn Y}
 \global\long\def\bV{\bn V}
 \global\long\def\bW{\bn W}
 \global\long\def\bG{\bn G}

\global\long\def\bM{\bn M}
 \global\long\def\bm{\bn m}
 \global\long\def\bE{\bn E}
 \global\long\def\bK{\bn K}
 \global\long\def\bL{\bn L}
 \global\long\def\bC{\bn C}
 \global\long\def\bT{\bn T}
 \global\long\def\bg{\bn g}
 \global\long\def\bN{\bn N}
 \global\long\def\bX{\bn X}
 \global\long\def\bR{\bn R}
 \global\long\def\bHs{\bn{Hs}}
 \global\long\def\bDelta{\bn{\Delta}}
 \global\long\def\bsigma{\bn{\sigma}}
 \global\long\def\bpsi{\bn{\psi}}
 \global\long\def\bQ{\bn Q}
 \global\long\def\bZ{\bn Z}
 \global\long\def\bU{\bn U}
 \global\long\def\bz{\bn z}
 \global\long\def\bu{\bn u}
 \global\long\def\bk{\bn k}
 \global\long\def\bUpsilon{\bn{\Upsilon}}
 \global\long\def\bEta{\bn{\eta}}
 \global\long\def\bmu{\bn{\mu}}
 \global\long\def\brho{\bn{\rho}}
 \global\long\def\bpartial{\bn{\partial}}

\title{Dynamic homogenization of composite and locally resonant flexural
systems}

\author{René Pernas-Salomón and Gal Shmuel\thanks{Corresponding author. Tel.: +1 972 778871613. \emph{E-mail address}:
meshmuel@technion.ac.il (G. Shmuel).}\\
 {\small{}{}{}Faculty of Mechanical Engineering, Technion\textendash Israel
Institute of Technology, Haifa 32000, Israel}\\
 }
\maketitle
\begin{abstract}
Dynamic homogenization aims at describing the macroscopic characteristics
of wave propagation in microstructured systems. Using a simple method,
we derive frequency-dependent homogenized parameters that reproduce
the exact dispersion relations of infinitely periodic flexural systems.
Our scheme evades the need to calculate field variables at each point,
yet capable of recovering them, if wanted. Through reflected energy
analysis in scattering problems, we quantify the applicability of
the homogenized approximation. We show that at low frequencies, our
model replicates the transmission characteristics of semi-infinite
and finite periodic media. We quantify the decline in the approximation
as frequency increases, having certain characteristics sensitive to
microscale details. We observe that the homogenized model captures
the dynamic response of locally resonant media more accurately and
across a wider range of frequencies than the dynamic response of media
without local resonance. 

\end{abstract}

\section{Introduction}

The physics of systems with microstructure is governed by complex
differential equations with spatially varying coefficients, leading
to fields that exhibit rapid fluctuations at the microscale. Homogenization
theory aims at describing such systems in terms of simpler \emph{effective
}or\emph{ macroscopic} equations, assuming these microscale variations
can be averaged out \citep{Hashin1983,Nemat-Nasser1999,milt02book}.
In turn, the homogenized models\textemdash ordinarily developed when
analyzing infinite media\textemdash are employed in investigating
the physics of microstructured systems bordered by other media. Differently
from its constituents, the effective medium may exhibit extraordinary
properties, in which case it is termed a \emph{metamaterial}. Specifically,
metamaterials in elastodynamics admit negative effective mass and
stiffness, effective anisotropic mass, and capable of wave manipulation
through negative refraction, filtering and steering \citep{Milton07,bigoni2013prb,celli14,2016-acoustic-metamt-broad-horizons,Barnwell2017parnell}.

This work is concerned with the dynamic homogenization of composite
and locally resonant flexural media, whose dynamics and metamaterial
properties have been extensively studied recently \citep[\emph{e.g.},][]{2013-Xiao-periodic-beams-vibration-absorbers,carta15brun,CHEN2017179,Yang2017em}.
The Euler-Bernoulli beam model for flexural motions---the model we
address in the sequel---is not only one of the fundamental models
in structural engineering, it is also employed in MEMS modeling \citep{Korvink2006MEMS},
lattice models of materials \citep{Ostoja-Starzewski2002fk}, and
constitutes a platform for the analysis of novel applications \citep{Colquitt2014,Misseroni2016sr,CHEN2017179,Zareei2018}.
Different approaches were employed to describe their effective behavior;
\citet{2017-Sun-IJAppMech} used an asymptotic expansion method to
derive effective wave equations at low frequencies of a composite
beam, rather than identifying effective properties of a homogenized
medium; \citet{2012-High-frequency-plates-FWaves} extended the exceptional
high frequency homogenization theory of \citet{2010-Craster-PRSA}
to obtain a long-scale governing equation\textemdash of a different
form than the microscale equation\textemdash that is applicable at
high frequencies, by perturbing about standing long-waves; \citet{CHEN2017179}
defined frequency-dependent effective properties of a beam with periodically
attached local resonators in terms of calculated macroscopic quantities;
\citet{torrent2014prb} developed an effective theory for inclusion-based
locally resonant flexural media, based on the scattering properties
of the inclusions.

The objective of the present work is twofold; (\emph{i}) Develop a
simple homogenization scheme that delivers the exact dispersion relation---not
only at low frequencies---for composite beams and flexural systems
with periodically attached local resonators; (\emph{ii}) Quantify
the implications of violating the fundamental homogenization assumption\textemdash that
the wavelength is much larger than the microstructure\textemdash on
the replacement of semi-infinite and finite systems by their homogenized
models. 

To achieve objective (\emph{i}), we examine macroscopic equations
for the volume averages of the field variables. The corresponding
effective coefficients are derived from micromechanical considerations
and Fourier analysis, without the need to calculate the field variables
at each point \citep[\emph{cf}., ][]{Willis2009,NN11srivastava}.
A similar approach was applied by \citet{NematNasser2011jmps} for
dynamic homogenization of laminates; as in the latter work, the local
fields are actually extractable from our scheme, if wanted. Indeed,
the macroscopic equations, in conjunction with the frequency-dependent
effective properties, deliver the exact dispersion relation of infinite
microstructured media. 

To carry out objective (\emph{ii}), we investigate the reflection
behavior of semi-infinite and finite periodic systems in comparison
with their homogenized replacements, through a study of two scattering
problems, as investigated by \citet{Srivastava2014} and \citet{joseph2015}
for laminates. In the first problem, we analyze the energy reflected
from an interface between two semi-infinite media, where one is periodic
and the other is its homogenized equivalent. Specifically, we explore
the reflected energy dependency on the frequency and microscale details,
such as the interface location within the periodic cell. In the second
problem, we analyze a finite periodic beam bounded between two semi-infinite
homogeneous beams, and compare its transmission spectrum with the
homogenized equivalent spectrum. 

The paper is organized as follows. Sec. \ref{Homogenization-scheme}
firstly revisits the problem flexural wave propagation in composite
beams and systems with local resonators. Afterwards, our derivation
for the macroscopic equations and effective properties is provided.
 In Sec.\;\ref{Quantifying-applicability} we describe the semi-infinite
and finite scattering problems, and derive expressions for the reflected
and transmitted energy, respectively. Therein, we demonstrate that
in the long-wavelength limit, our homogenized model is able to match
the impedance of the original periodic system, and hence to avoid
reflection. Sec. \ref{Numerical-investigation} studies the applicability
of the homogenization model in the infinite problem and scattering
problems by way of numerical examples. Sec. \ref{Conclusion} concludes
the paper, summarizing our main results and observations.

\section{\label{Homogenization-scheme}Dynamic homogenization for periodic
flexural systems }


\subsection{Wave propagation in periodic flexural systems}

\label{sec-periodic-system}

\begin{figure}[t!]
\centering \includegraphics[width=1\textwidth]{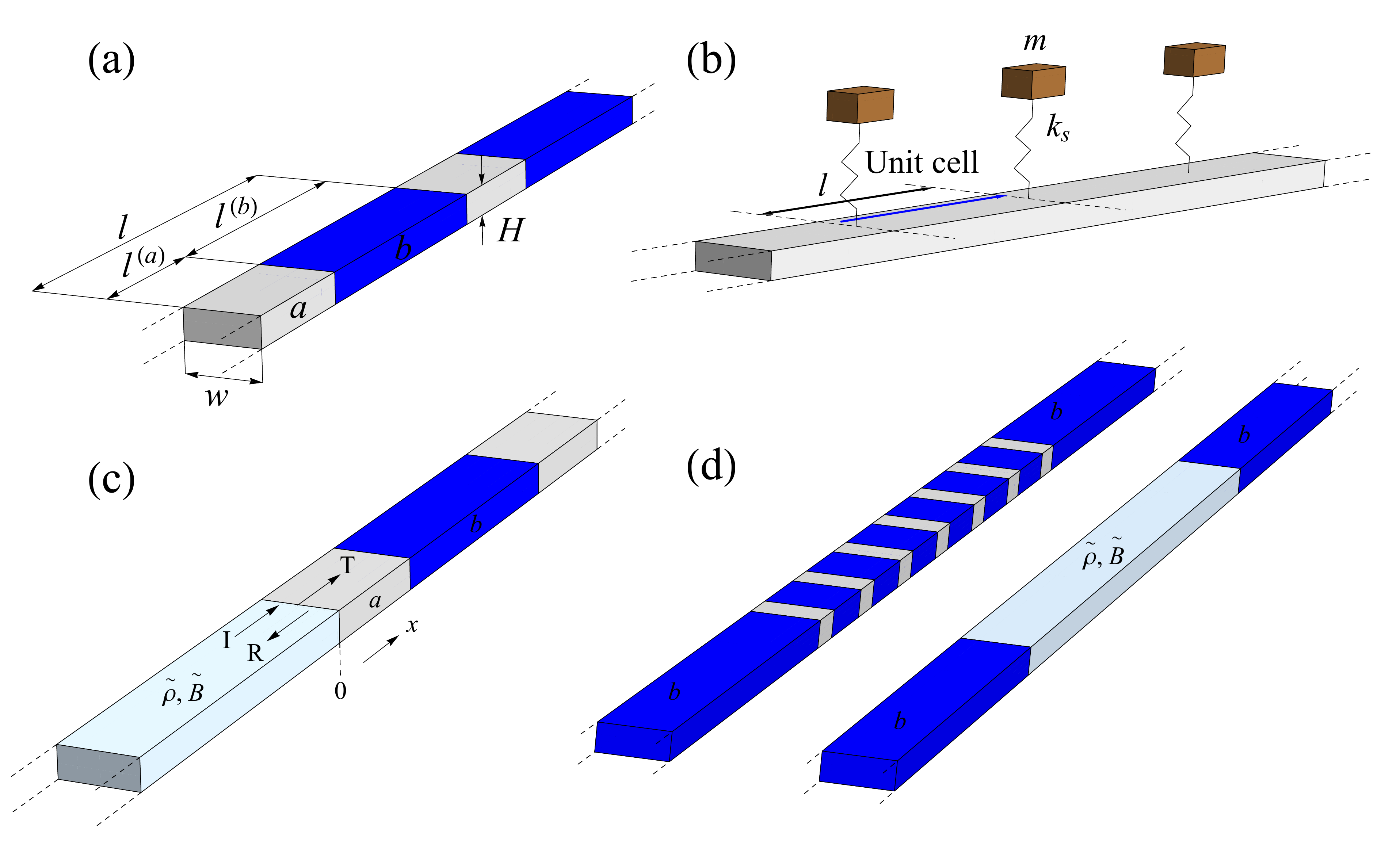}
\caption{\label{Fig-1} (a) An infinite periodic beam made of two alternating
$a$ and $b$ phases.  (b) A flexural system with periodically attached
local resonators, modeled as mass-spring elements attached to a uniform
beam. A unit cell of length $l$ comprises a resonator attached to
its right end. (c) A semi-infinite periodic beam in contact at $x=0$
with its homogenized equivalent occupying the domain $x<0$. The letters
I, R, and T denote incident, reflected and transmitted waves, respectively.
(d) Two semi-infinite $b$ phases connected by a finite periodic beam
(left), and connected by the homogenized beam of the same length (right).}
\end{figure}

\emph{Composite beams.} Consider a beam made of alternating phases
in the $x$ direction, namely, phases $a$ and $b$ of lengths $l^{\left(a\right)}$
and $l^{\left(b\right)}$, respectively. Accordingly, we have that
$E(x+l)=E(x)$ and $\rho(x+l)=\rho(x)$, where $l=l^{\left(a\right)}+l^{\left(b\right)}$
is the length of the unit cell, $E(x)$ is the Young modulus and $\rho(x)$
the mass density per unit volume. The cross-section area, $A$, and
the inertia moment, $I$, are uniform throughout the medium, as illustrated
in Fig.\;\ref{Fig-1}(a).


In the absence of distributed loading, the Euler-Bernoulli beam model
of flexural motion reads \citep[see, \emph{e.g.},][]{graff1975wave}
\begin{eqnarray}\label{Eq-flexural-motion} 
\left[ B(x)u_{,xx}\right]{_{,xx}}+ \rho(x) A u_{,tt} &=& 0, \end{eqnarray}where $u$ is the transverse displacement and $B(x)=E(x)I$ is the
bending stiffness. Assuming time dependency in the form $e^{-i\omega t}$,
Eq. (\ref{Eq-flexural-motion}) can be rewritten as \begin{eqnarray}\label{GE} M_{,xx} - i\omega p &=&0, \end{eqnarray} 
$M=B(x)\,u_{,xx}$ and 
\begin{eqnarray}
p & = & \rho(x)Au_{,t}=-\rho(x)Ai\omega u\label{Constitutive-relations-Vd}
\end{eqnarray}
is the linear momentum. For harmonic waves traveling in this periodic
beam, the field variables are of the form \citep{1928-Bloch-Zeit-fur-Physik,Kittel}
\begin{eqnarray}\label{Eq-1} R(x,t) &=& R_\mathrm{p}(x)e^{i (\Blochwn x-\omega t)}, \end{eqnarray} where
$R$ represents the field variables, namely, $u$, and the angle of
rotation, $\theta$, bending moment, $M$ and shear force, $V$, and
$R_{\mathrm{p}}$ is periodic with the same periodicity as the unit
cell; the outstanding question is to relate the Bloch wavenumber,
$\Blochwn$, and the frequency $\omega$. To this end, we note that
in each phase, the general solution of Eq. (\ref{Eq-flexural-motion})
is 
\begin{eqnarray}
u(x)=C^{+}e^{i\waven x}+D^{+}e^{-\waven x}+C^{-}e^{-i\waven x}+D^{-}e^{\waven x},
\end{eqnarray}
where $\waven=\sqrt[4]{\rho A\omega^{2}/B}$, and the coefficients
$C^{\pm}$ (resp. $D^{\pm}$) denote the amplitudes of the propagating
(resp. non-propagating) waves. It follows that the field variables
at the ends $x_{0}^{(i)}$ and $x_{0}^{(i)}+l^{(i)}$ of each phase
$i$ are related via\footnote{The standard transfer matrix formulation is given, \emph{e.g.}, in
\citet{carta15brun} and \citet{XU2016}. } 
\begin{eqnarray}
\left\lbrace \begin{array}{c}
u\left(x_{0}^{(i)}\right)\\
\theta\left(x_{0}^{(i)}\right)\waven^{(a)^{-1}}\\
M\left(x_{0}^{(i)}+l^{(i)}\right)\waven^{(a)^{-2}}\\
V\left(x_{0}^{(i)}+l^{(i)}\right)\waven^{(a)^{-3}}
\end{array}\right\rbrace =\mathsf{H}^{(i)}\cdot\left\lbrace \begin{array}{c}
M\left(x_{0}^{(i)}\right)\waven^{(a)^{-2}}\\
V\left(x_{0}^{(i)}\right)\waven^{(a)^{-3}}\\
u\left(x_{0}^{(i)}+l^{(i)}\right)\\
\theta\left(x_{0}^{(i)}+l^{(i)}\right)\waven^{(a)^{-1}}
\end{array}\right\rbrace ,\label{definition-Hyb-matrix}
\end{eqnarray}
where the $\waven^{(a)}$ is the value that $\waven$ takes in phase
$a$, and $\mathsf{H}^{(i)}$ is given in \nameref{Appendix-A}. Using
the continuity of the field variables and Eq.\;(\ref{Eq-1}), the
dispersion relation $\Blochwn(\omega)$ of a periodic beam can be
determined by solving the generalized eigenproblem \citep{2010-Tan} 

\begin{eqnarray}
\label{EP-1} 
\left[ \begin{array}{cc} -\mathsf{I}_{2} & \mathsf{H}_{11} 
\\ \mathsf{0}_2 & \mathsf{H}_{21} \\ \end{array} \right]\cdot\mathsf{s}_\mathrm{m}(x)  &=& e^{i \Blochwn l}\left[ \begin{array}{cc} -\mathsf{H}_{12} &  \mathsf{0}_2 \\ -\mathsf{H}_{22} & \mathsf{I}_{2} 
\\ 
\end{array} \right]\cdot \mathsf{s}_\mathrm{m}(x);
\end{eqnarray} here $\mathsf{H}_{11},\mathsf{H}_{12},\mathsf{H}_{21},\mathsf{H}_{22}$
denote the $2\times2$ sub-blocks of the hybrid matrix corresponding
to the unit cell whose ends are at $x$ and $x+l$, and $\mathsf{s}_{\mathrm{m}}(x)=\left\lbrace u(x),\theta(x)\waven^{(a)^{-1}},M(x)\waven^{(a)^{-2}},V(x)\waven^{(a)^{-3}}\right\rbrace ^{\mathsf{T}}$
is the modified \textit{state vector}. The characteristic polynomial
associated with Eq.\;(\ref{EP-1}) provides 
\begin{eqnarray}
\cos\Blochwn l & = & \dfrac{-a_{2}\pm\sqrt{a_{2}^{2}-4a_{1}\left(a_{3}-2a_{1}\right)}}{4a_{1}},\label{Closed-E}
\end{eqnarray}
where $a_{1}=\mathrm{det}\,\mathsf{H}_{12}$ and the coefficients
$a_{2}$ and $a_{3}$ are cumbersome functions of the matrix elements
in Eq.\;(\ref{EP-1}), omitted for brevity.

Note that if $\Blochwn l$ is a solution, then so are $2\pi n\pm\Blochwn l$,
for $n\in\mathbb{Z}$. The region $-\pi\leqslant\Blochwn l\leqslant\pi$
is called the $1^{\mathrm{st}}$ Brillouin zone. The $2^{\mathrm{nd}}$
Brillouin zone comprises the negative region $-2\pi+\Blochwn l$ and
the positive region $2\pi-\Blochwn l$, and so forth. To determine
the range of propagating and attenuating frequencies it is sufficient
to examine the irreducible $1^{\mathrm{st}}$ Brillouin zone, $0\leqslant\Blochwn l\leqslant\pi$
\citep{Farhad-2011}; if $\Blochwn$ is complex, then the frequency
belongs to a gap, \emph{i.e.}, there is no propagating solution, and
waves at this frequency decay.

\emph{Locally resonant flexural systems}. The dispersion relation
derivation of flexural systems with periodically attached local resonators
is summarized next \citep{Dianlong2006,2013-Xiao-periodic-beams-vibration-absorbers,Shuguang2015,carta15brun}.
Such systems are analyzed as uniform Euler-Bernoulli beams connected
periodically to harmonic oscillators modeled as mass\,($m$)-spring\,($k_{s}$)
elements, as shown in Fig.\;\ref{Fig-1}(b). The equation of motion
for this model can be expressed as 
\begin{eqnarray}
\left[B(x)\,u_{,xx}\right]{_{,xx}}-\rho(x)\omega^{2}A\,u(x) & = & -\sum_{n}\frac{k_{s}\omega^{2}\,u(x_{n})}{\omega^{2}-\omega_{0}^{2}}\delta(x-x_{n}),\label{Equation-original-spring}
\end{eqnarray}
where $\delta(x)$ is the Dirac delta function, $\omega_{0}=\sqrt{\dfrac{k_{s}}{m}}$
and the factor $-\dfrac{k_{s}\omega^{2}u(x_{n})}{(\omega^{2}-\omega_{0}^{2})}$
is the force acting on the beam at the connection points $x_{n}=nl$,
$n\in\mathbb{Z}$. Note that $\sum_{n}u(x_{n})\delta(x-x_{n})=u(x)\sum_{n}\delta(x-x_{n})$,
hence we can recast Eq. (\ref{Equation-original-spring}) in a form
that is similar to Eq.\;(\ref{Eq-flexural-motion}), by replacing
$\rho(x)$ with $\hat{\rho}(x)=\rho(x)-\dfrac{k_{s}}{(\omega^{2}-\omega_{0}^{2})A}\sum_{n}\delta(x-x_{n})$.

As in the previous derivation, a hybrid matrix and Bloch-Floquet analysis
is carried out by considering a unit cell of length $l$ with a resonator
attached to its right end. The resultant hybrid matrix is equal to
the hybrid matrix of a homogeneous cell plus the term $\dfrac{k_{s}\omega^{2}}{\left(\omega^{2}-\omega_{0}^{2}\right)}$
at its $\left(4,3\right)$ entry (\nameref{Appendix-A}). The appended
term comes from the change in the value of the shear force at the
connection point. The corresponding dispersion relation is Eq. (\ref{Closed-E})
with modified $a_{i}$ according to the new hybrid matrix. 

\subsection{Effective properties}

\label{sec-Effective-properties}

Our derivation of the effective properties for flexural systems relies
on volume averages of field variables, similarly to the procedure
developed in \citet{NN11srivastava} for laminates \citep[see also][]{Willis2009}.
By construction, the resultant homogenized formulation satisfies macroscopic
field equations, and recovers exactly the dispersion relation of composite
and locally resonant beams.

To derive macroscopic relations in terms of mean quantities, we multiply
Eq. (\ref{GE}) by $e^{-i\Blochwn X}$ and obtain\begin{eqnarray}\label{GE-2} \left[M_\mathrm{p}(x)e^{i \Blochwn (x-X)}\right]{_{,xx}}- i\omega\, p_\mathrm{p} (x)e^{i \Blochwn (x-X)} &=&0. \end{eqnarray} In
terms of $\chi=x-X$, Eq. (\ref{GE-2}) reads \begin{eqnarray}\label{GE-3} \left[M_\mathrm{p}(X+\chi)e^{i \Blochwn \chi}\right]{_{,\chi\chi}}- i\omega\, p_\mathrm{p} (X+\chi)e^{i \Blochwn \chi} &=& 0. \end{eqnarray} 
Eq. (\ref{GE-3}) with respect to $X$ over the unit cell provides
\begin{eqnarray}\label{GE-4} \left[\langle M_\mathrm{p}\rangle ^{i \Blochwn \chi}\right]{_{,\chi\chi}} - i\omega \langle p_\mathrm{p}\rangle e^{i \Blochwn \chi} &=&0, \end{eqnarray}where
the average of each one of the periodic parts is defined as 
\begin{eqnarray}
\left\langle \circ\right\rangle  & = & \frac{1}{l}\int_{-l/2}^{l/2}(\circ)\,\mathrm{d}X.
\end{eqnarray}
The averaged field variables are given by \begin{eqnarray}\label{Overal-Field-variables} \avevarsigma (x) &=&  \langle \varsigma_\mathrm{p}\rangle e^{i \Blochwn x},\;\;  \varsigma=M, p,u, \end{eqnarray}and
satisfy a governing equation similar to Eq. (\ref{GE}), in the form
\begin{eqnarray}\label{Eq-overall} \aveM (x)_{,xx} - i\omega \avep (x) &=&0. \end{eqnarray}We
complete our formulation with the following macroscopic counterparts
of the moment-displacement relation and Eq. (\ref{Constitutive-relations-Vd})
\begin{eqnarray}\label{Effective-parameters-definition} \aveM (x) &=& \Beff\, \aveu_{,xx}(x),\;\;\; \avep (x)= -i\omega A \rhoeff\, \aveu(x), \end{eqnarray}which
define the effective bending stiffness, $\Beff$, and effective mass
density, $\rhoeff$. Eq. (\ref{Effective-parameters-definition})
yields with Eq. (\ref{Eq-overall}) a frequency-wavenumber relation
whose form is analogous to that of a homogeneous beam, namely, \begin{eqnarray}\label{DRelation} \Blochwn^4-\omega^2\frac{A\, \rhoeff}{\Beff} &=& 0.  \end{eqnarray}Eq.
(\ref{DRelation}) reproduces precisely the dispersion relation of
flexural composite and locally resonant systems, through the frequency-dependent
effective properties $\rhoeff$ and $\Beff$ given in Eq. \eqref{Eq-overall},
as will be demonstrated in the sequel.

\subsection{Analytic formulas for calculating the effective properties}

\noindent \label{Effective-parameters} The common procedure to determine
the effective properties requires the calculation of the local fields
and integration of their periodic parts over the unit cell. Here we
employ a different approach, incorporating Fourier analysis. Firstly,
we consider the Bloch form of $u,M$ and $p$, and expand their periodic
part into Fourier series, as well as the quantities $B$ and $\rho$
(or $\hat{\rho}$ for the locally resonant beam). Accordingly, we
have that \begin{eqnarray}\label{FE-varsigma} \varsigma(x) &=& e^{i\Blochwn x} \sum_{m}\varsigma_\mathrm{F}(m) e^{\frac{2 i \pi m x}{l}},\;\;\;\;\;  \varsigma=u, M, p,\;\;\;\;\;m\in \mathbb{Z}, \end{eqnarray}where
\begin{eqnarray} \varsigma_\mathrm{F}(m) &=& \frac{1}{l} \int_{-l/2}^{l/2}  \varsigma_\mathrm{p}(x) e^{\frac{-2 i \pi m x}{l}} \mathrm{d}x. \end{eqnarray}In
Eq. (\ref{FE-varsigma}), the term with $m=0$ describes the part
of $\varsigma(x)$ that varies slower with $x$ than the part associated
with the rest of the terms. It follows that \begin{eqnarray}\label{varsigma0} \avevarsigma(x) &=& \varsigma_\mathrm{F}(m=0)e^{i\Blochwn x} =\langle \varsigma_\mathrm{p}\rangle e^{i\Blochwn x}. \end{eqnarray}Substitution
of the Fourier expansions into the relation $M=B(x)\,u_{,xx}$ and
Eq. (\ref{Constitutive-relations-Vd}) delivers a relation between
$\langle M_{\mathrm{p}}\rangle$ and $\langle u_{\mathrm{p}}\rangle$,
and between $\langle p_{\mathrm{p}}\rangle$ and $\langle u_{\mathrm{p}}\rangle$,
namely, %

\noindent \begin{flalign} 
&\langle M_\mathrm{p}\rangle =  -\BF(0) \Blochwn^2 \langle u_\mathrm{p}\rangle -  \sum_{m\neq0} \BF(-m) \left(\Blochwn+\frac{2 \pi m }{l}\right)^2 \uF(m),  \label{M0-um} \\ 
&\langle p_\mathrm{p}\rangle = -\rhoF(0) i\omega A \langle u_\mathrm{p}\rangle -i\omega A\,\sum_{m\neq0} \rhoF(-m)\,\uF(m).  \label{p0-expanded} \end{flalign}We clarify that in Eq.\;(\ref{p0-expanded}) and the sequel, the
terms $\rhoF$ are replaced by $\hat{\rho}_{\mathrm{F}}$ when the
locally resonant beam is addressed. We eliminate $\uF(m)$ from Eqs.
(\ref{M0-um}-\ref{p0-expanded}) by expressing it in terms of $\langle u_{\mathrm{p}}\rangle$,
similarly to \citet{2009-Felipe-GetPDFServlet}, who carried out a
related analysis for photonic crystals. In our case, this relation
is obtained from the equation describing the propagation of flexural
waves in beams, \emph{i.e.}, Eq. (\ref{Eq-flexural-motion}). Substituting
into this equation the Fourier expansions provides \begin{eqnarray}\label{Secular-Equation} \sum_{m^{\prime}}   \mathsf{Q} (\Blochwn; m,m^{\prime}) \uF(m^{\prime})   &=& 0, \end{eqnarray}where
\begin{eqnarray}\label{Q-app} \mathsf{Q} (\Blochwn; m,m^{\prime}) &=& \left(\Blochwn+\frac{2 \pi m}{l}\right)^2 \BF (m-m^{\prime}) \left(\Blochwn+\frac{2 \pi m^{\prime}}{l}\right)^2 - \omega^2 A\, \rhoF(m-m^{\prime}). \end{eqnarray}We
write the coefficients $\uF(m\neq0)$ in terms of $\uF(m=0)=\langle u_{\mathrm{p}}\rangle$
via the equations for $m\neq0$ in Eq. (\ref{Secular-Equation}),
and obtain \begin{eqnarray}
\label{um-app} 
\uF(m\neq0) &=& -\sum_{m^{\prime}\neq0} \mathsf{Q}_s^{-1} (\Blochwn; m,m^{\prime}) \mathsf{Q} (\Blochwn; m^{\prime},0) \langle u_\mathrm{p}\rangle. \end{eqnarray}Here, $\mathsf{Q}_{s}(\Blochwn;m,m^{\prime})$ is a sub-matrix, obtained
from the matrix represented in Eq. (\ref{Q-app}) after eliminating
its row (resp. column) for $m=0$ (resp. $m^{\prime}=0$). The effective
properties $\Beff$ and $\rhoeff$ are determined by substituting
Eq. (\ref{um-app}) into Eqs. (\ref{M0-um}-\ref{p0-expanded}) and
utilizing the macroscopic relations $\langle M_{\mathrm{p}}\rangle=-\Beff\,\Blochwn^{2}\langle u_{\mathrm{p}}\rangle$
and $\langle p_{\mathrm{p}}\rangle=-i\omega A\rhoeff\langle u_{\mathrm{p}}\rangle$.
The end result reads 
\begin{eqnarray}
\begin{aligned} & \Beff(\Blochwn,\omega)=\BF(0)+\sum_{m\neq0}\BF(-m)\left(\Blochwn+\frac{2\pi m}{l}\right)^{2}\wF(m),\\
 & \wF(m\neq0)=-\sum_{m^{\prime}\neq0}\mathsf{Q}_{s}^{-1}(\Blochwn;m,m^{\prime})\left[\left(\Blochwn+\frac{2\pi m^{\prime}}{l}\right)^{2}\BF(m^{\prime})-\left(\frac{\omega}{\Blochwn}\right)^{2}A\rhoF(m^{\prime})\right].
\end{aligned}
\label{Beff-1}
\end{eqnarray}
\begin{equation}
\begin{aligned} & \rhoeff(\Blochwn,\omega)=\rhoF(0)+\sum_{m\neq0}\rhoF(-m)v_{\mathrm{F}}(m),\\
 & v_{\mathrm{F}}(m\neq0)=-\sum_{m^{\prime}\neq0}\mathsf{Q}_{s}^{-1}(\Blochwn;m,m^{\prime})\left[\left(\Blochwn+\frac{2\pi m^{\prime}}{l}\right)^{2}\BF(m^{\prime})\,\Blochwn^{2}-\omega^{2}A\rhoF(m^{\prime})\right].
\end{aligned}
\label{Rhoeff}
\end{equation}
We remark that for the locally resonant beam, the components $\hat{\rho}_{\mathrm{F}}(m-m^{\prime})$
are negative at ${1<\dfrac{\omega}{\omega_{0}}<\sqrt{1+\dfrac{m}{\rho Al}}}$.
Notably, the effective properties $\Beff$ and $\rhoeff$ depend on
the wavenumber $\Blochwn$ and the frequency $\omega$. Eq.\;(\ref{Secular-Equation})
provides the dispersion relation $\Blochwn(\omega)$ of the flexural
system, through its implied condition 
\begin{eqnarray}
\mathrm{det}\,\mathsf{Q}=0.
\end{eqnarray}
(Of course, this calculation of the dispersion relation is not needed,
having Eq. \eqref{Closed-E} at hand; we provide it to argue that
our approach can be generalized to cases in which exact dispersion
relations are not accessible.) In turn, the frequency-dependent effective
properties $\Beff$ and $\rhoeff$ are obtained by substituting ($\Blochwn,\omega$)
pairs into Eqs.\;(\ref{Beff-1}-\ref{Rhoeff}). Alternatively, the
dispersion relation and the effective properties can be evaluated
using the following iterative procedure. Firstly, the static weighted
averages 
\begin{eqnarray}
\Beff_{0}=l\left(\dfrac{l^{(a)}}{B^{(a)}}+\dfrac{l^{(b)}}{B^{(b)}}\right)^{-1},\;\rhoeff_{0}=\frac{l^{(a)}\rho^{(a)}+l^{(b)}\rho^{(b)}}{l}
\end{eqnarray}
are substituted into Eq. (\ref{DRelation}) as $\Beff$ and $\rhoeff$
to obtain an initial dispersion relation ($\Blochwn,\omega$); the
latter is substituted into Eqs.\;(\ref{Beff-1}-\ref{Rhoeff}) to
find a first iteration of the frequency-dependent properties $\Beff(\Blochwn,\omega)$
and $\rhoeff(\Blochwn,\omega)$, which are substituted back to Eq.
(\ref{DRelation}) to evaluate the next iteration of the dispersion
relation, and so forth, until convergence. 

We emphasize that our scheme evades the need to calculate the microscopic
displacement field; it is actually extractable from our scheme, by
substituting back $\mathsf{Q}$ into Eq.\;(\ref{Secular-Equation})
to calculate $\uF(m)$, and in turn, $u(x)$.

\section{\label{Quantifying-applicability}Reflection and transmission in
interface problems}

As discussed by \citet{Srivastava2014} and \citet{joseph2015}, to
justify a replacement of the periodic medium by fictitious homogeneous
medium with effective properties, their response to interface problems
should be similar. \citet{Srivastava2014} suggested to quantify this
similarity by the reflected energy \citep[see also][]{2014-Effective-medium-breakdown-Hanan-Segev,joseph2015,AMIRKHIZI2017}.
Accordingly, we analyze next the reflection and transmission at the
interface of semi-infinite and finite periodic flexural systems, in
comparison with their dynamic homogenized equivalents, as illustrated
in Figs.\,\ref{Fig-1}(c) and \ref{Fig-1}(d). 

\subsection{Two contiguous semi-infinite beams }

\label{Reflection-problem} We consider a semi-infinite fictitious
homogeneous beam with bending stiffness $\Beff$ and mass density
$\rhoeff$ occupying the domain $x<0$. The beam is perfectly bounded
at $x=0$ to a semi-infinite periodic system, whose properties are
described in Sec.\;\ref{sec-periodic-system}, occupying the domain
$x>0$, see Fig.\,\ref{Fig-1}(c). The fictitious beam serves as
the homogenized equivalent of the periodic system, hence its properties
satisfy the dispersion relation (\ref{DRelation}), with a wavenumber
that coincides with the Bloch wavenumber $\Blochwn$ of waves in the
periodic system. At the interface, a positive-going wave $C^{+}e^{i\Blochwn x}$
excited from the left is partially reflected in the form of a negative-going
wave $C^{-}e^{-i\Blochwn x}$ and attenuating wave $D^{-}e^{\Blochwn x}$.
A transmitted wave in the semi-infinite periodic domain is created
too, comprising positive-going wave $C_{\mathrm{p}}^{+}u_{\mathrm{p}}(x)e^{i\Blochwn x}$
and attenuating wave $D^{+}e^{-\Blochwn x}$. The attenuating waves\textemdash required
for the field variables to be continuous across the interface\textemdash are
negligible far from the interface. From the corresponding continuity
conditions at the interface we obtain \begin{eqnarray}\label{CR-2} \begin{aligned} &C^+ + C^- + D^-= C_\mathrm{p}^{+} u(0)+ D^+, \\ &i C^+ -i C^- + D^-= C_\mathrm{p}^{+} \frac{\theta(0)}{\Blochwn}- D^+, \\ &-\Beff C^+ - \Beff C^- + \Beff D^-= C_\mathrm{p}^{+} \frac{M(0)}{\Blochwn^2}+ \Beff D^+, \\ &i \Beff C^+ -i\Beff  C^- - \Beff D^-= C_\mathrm{p}^{+} \frac{V(0)}{\Blochwn^3}+ \Beff D^+, \end{aligned} \end{eqnarray}where
$u(0)$, $\theta(0)$, $M(0)$ and $V(0)$ are the solutions of the
hybrid matrix generalized eigenproblem (\ref{EP-1}) at $x=0$. Note
that the attenuating wave $D^{+}e^{-\Blochwn x}$ was treated as a
wave solution corresponding to a homogeneous medium having the dynamic
effective properties $\Beff$ and $\rhoeff$. The continuity conditions
(\ref{CR-2}) constitute a system of linear equations, from which
we obtain the reflection coefficient $r\equiv\dfrac{C^{-}}{C^{+}}$,
namely, 
\begin{eqnarray}
r & = & \frac{i\,\Blochwn^{3}\Beff u(0)-i\,\Blochwn M(0)-\left[V(0)+\Blochwn^{2}\Beff\theta(0)\right]}{i\,\Blochwn^{3}\Beff u(0)-i\Blochwn M(0)+\left[V(0)+\Blochwn^{2}\Beff\theta(0)\right]},\label{r-app-2}
\end{eqnarray}
or \begin{eqnarray}\label{NRE} r &=& \frac{\gamma-1}{\gamma+1}, \end{eqnarray}where
\begin{eqnarray}
\gamma & = & \frac{i\Blochwn^{3}\,\Beff\,u(0)-i\Blochwn M(0)}{V(0)+\Blochwn^{2}\,\Beff\,\theta(0)}.\label{Gamma}
\end{eqnarray}
The normalized reflected energy equals $\left|r\right|^{2}$, and
vanishes when $\gamma=1$. Following \citet{Srivastava2014} , we
use it to quantify the applicability of the homogenized models, in
our context of flexural systems. Furthermore, bearing in mind that
both the reflective energy and the effective properties depend on
the chosen solution of $\Blochwn$, \citet{Srivastava2014} suggested
to use energy conservation requirements to determine which solution
should be used for $r$, and, in turn, calculating $\rhoeff$ and
$\Beff$. This is demonstrated in the sequel.

\emph{Long-wavelength limit and reflection dependency on the interface}
\emph{location}. It is expected that in the long-wavelength limit
$\gamma\approx1$ and therefore $\left|r\right|^{2}\approx0$. We
now verify that our derivation for the effective properties meets
this expectation. Since $e^{i\Blochwn x}=1$ at $x=0^{+}$, we have
that \begin{eqnarray} u(0) = u_\mathrm{p}(0),\;\;\theta(0) = \theta_\mathrm{p}(0),\;M(0) = M_\mathrm{p}(0),\;V(0) = V_\mathrm{p}(0). \end{eqnarray}In
the long-wavelength limit, the periodic part of the displacement varies
slowly over the unit cell, its derivatives practically vanish and
we have $\theta_{\mathrm{p}}(0)\approx i\Blochwn u_{\mathrm{p}}(0)$,
$V_{\mathrm{p}}(0)\approx-i\Blochwn M_{\mathrm{p}}(0)$. Then, the
parameter $\gamma$ becomes 
\begin{eqnarray}
\gamma\approx\frac{i\Blochwn^{3}\,\Beff\,u_{\mathrm{p}}(0)-i\Blochwn M_{\mathrm{p}}(0)}{-i\Blochwn M_{\mathrm{p}}(0)+\Blochwn^{2}\,\Beff\,i\Blochwn u_{\mathrm{p}}(0)}=1.\label{gamma-low-freq}
\end{eqnarray}
We have that $\gamma\approx1$ and, in turn, $r\approx0$ as it should.
Note that this result is only satisfied for a homogeneous beam having
the effective properties given in Eq.\;(\ref{Effective-parameters-definition}).
If this medium is replaced by a homogeneous one having mass density
$\rho_{h}\neq\rhoeff$ and bending stiffness $B_{h}\neq\Beff$ (which
also satisfy the dispersion relation of the periodic beam), then Eqs.\;(\ref{CR-2})
will not yield vanishing $r$ in the low frequency limit. Stated differently,
in the long-wavelength limit the periodic structure effectively behaves
as a homogeneous beam whose bending stiffness is $\Beff\approx-M_{\mathrm{p}}(0)/\Blochwn^{2}u_{\mathrm{p}}(0)$. 

Note that the result in Eq. (\ref{gamma-low-freq}) is independent
of the interface location within the unit cell. Generally, however,
the reflected energy depends on that location, and hence, so does
the applicability of the homogenized model. This is due to the dependency
of the field variables on the position, and in turn, the parameter
$\gamma$, as we numerically demonstrate in Sec. \ref{Numerical-investigation}. 

\subsection{Finite system bounded by two semi-infinite homogeneous media}

We analyze the transmission through a periodic system bounded by two
homogeneous semi-infinite media, see Fig.\,\ref{Fig-1}(d). Specifically,
we consider an incident wave of amplitude $C^{+}(L)$ from the left,
and calculate the normalized transmitted energy, $\left|t\right|^{2}=\left|\dfrac{C^{+}(R)}{C^{+}(L)}\right|^{2}$,
where the $C^{+}(R)$ is the magnitude of the transmitted wave to
the right. To evaluate $\left|t\right|^{2}$, we derived an expression
using the hybrid matrix of the intermediate finite system, $\bH_{\mathrm{f}}$,
determined according to the procedure detailed in \nameref{Appendix-A}.
For simplicity, we assume that the semi-infinite media are made of
phase $b$. The corresponding components of the modified state vector
$\mathsf{s}_{\mathrm{m}}(x)$ are 
\begin{eqnarray}
\begin{aligned} & \left\lbrace \begin{array}{c}
u(x)\\
\theta(x)\waven^{(a)^{-1}}
\end{array}\right\rbrace =\mathsf{Q}_{u\theta}^{+}(x)\cdot\mathsf{c}^{+}+\mathsf{Q}_{u\theta}^{-}(x)\cdot\mathsf{c}^{-},\\
 & \left\lbrace \begin{array}{c}
M(x)\waven^{(a)^{-2}}\\
V(x)\waven^{(a)^{-3}}
\end{array}\right\rbrace =\mathsf{Q}_{MV}^{+}(x)\cdot\mathsf{c}^{+}+\mathsf{Q}_{MV}^{-}(x)\cdot\mathsf{c}^{-},
\end{aligned}
\end{eqnarray}
where 
\begin{eqnarray}
\begin{aligned} & \mathsf{c}^{\pm}=\left\lbrace \begin{array}{c}
C^{\pm}\\
D^{\pm}
\end{array}\right\rbrace ,\;\mathsf{Q}_{u\theta}^{\pm}(x)=\left[\begin{array}{cc}
e^{\pm i\waven^{(b)}x} & e^{\mp\waven^{(b)}x}\\
\pm i\dfrac{\waven^{(b)}}{\waven^{(a)}}e^{\pm i\waven^{(b)}x} & \mp\dfrac{\waven^{(b)}}{\waven^{(a)}}e^{\mp\waven^{(b)}x}
\end{array}\right],\\[5pt]
 & \mathsf{Q}_{MV}^{\pm}(x)=B^{(b)}\left(\dfrac{\waven^{(b)}}{\waven^{(a)}}\right)^{2}\left[\begin{array}{cc}
-e^{\pm i\waven^{(b)}x} & e^{\mp\waven^{(b)}x}\\
\pm i\dfrac{\waven^{(b)}}{\waven^{(a)}}e^{\pm i\waven^{(b)}x} & \pm\dfrac{\waven^{(b)}}{\waven^{(a)}}e^{\mp\waven^{(b)}x}
\end{array}\right].
\end{aligned}
\label{coefficients}
\end{eqnarray}
Here, $B^{(b)}$ and $\waven^{(b)}$ denote, respectively, the bending
stiffness and the wavenumber in phase $b$. The scattering matrix
$\mathsf{S}$ 
\begin{equation}
\begin{array}{ccc}
\mathsf{S} & = & \left[\left[\begin{array}{cc}
\mathsf{Q}_{u\theta}^{-}(0) & \bcero_{2}\\
\bcero_{2} & \mathsf{Q}_{MV}^{+}(0)
\end{array}\right]-\bH_{\mathrm{f}}\cdot\left[\begin{array}{cc}
\mathsf{Q}_{MV}^{-}(0) & \bcero_{2}\\
\bcero_{2} & \mathsf{Q}_{u\theta}^{+}(0)
\end{array}\right]\right]^{-1}\\
 &  & \cdot\left[\bH_{\mathrm{f}}\cdot\left[\begin{array}{cc}
\mathsf{Q}_{MV}^{+}(0) & \bcero_{2}\\
\bcero_{2} & \mathsf{Q}_{u\theta}^{-}(0)
\end{array}\right]-\left[\begin{array}{cc}
\mathsf{Q}_{u\theta}^{+}(0) & \bcero_{2}\\
\bcero_{2} & \mathsf{Q}_{MV}^{-}(0)
\end{array}\right]\right]
\end{array}\label{eq:S1M def}
\end{equation}
($\bcero_{2}$ is the $2\times2$ null matrix) relates the amplitudes
at the semi-infinite media, namely, %
\begin{eqnarray}
\left\lbrace \begin{array}{c}
\mathsf{c}^{-}(\mathrm{L})\\
\mathsf{c}^{+}(\mathrm{R})
\end{array}\right\rbrace =\mathsf{S}\cdot\left\lbrace \begin{array}{c}
\mathsf{c}^{+}(\mathrm{L})\\
\mathsf{c}^{-}(\mathrm{R})
\end{array}\right\rbrace .
\end{eqnarray}
In terms of $\mathsf{S}$, the normalized transmitted energy is 
\begin{eqnarray}
\left|t\right|^{2}=\left|\left\lbrace 1\;\;0\right\rbrace \cdot\mathsf{S}_{21}\cdot\left\lbrace \begin{array}{c}
1\\
0
\end{array}\right\rbrace \right|^{2},\label{transs-t}
\end{eqnarray}
where $\mathsf{S}_{21}$ is the $2\times2$ bottom-left block of $\mathsf{S}$.
Expression (\ref{transs-t}) is also used to evaluate the normalized
transmitted energy when the intermediate finite system is a locally
resonant beam. In this case, the matrix $\bH_{\mathrm{f}}$ is based
on the modified hybrid matrix, described in Sec. \ref{sec-periodic-system}.

\section{\label{Numerical-investigation} Numerical calculations}

\noindent We quantify  next the applicability of our homogenization
scheme and study its dependency on the wavelength through numerical
realization of the previous derivations. This is carried out firstly
for composite media homogenization, and subsequently for locally resonant
media. 
\begin{table}[t!]
\centering %
\begin{tabular}{cccc}
\hline 
Phase \; \; & $\rho\,\mathrm{(kg/m^{3}})$ \; \; & $E$\,(Pa) \; \; & length\,(m)\tabularnewline
\hline 
$\mathit{a}$ \; \;  & 1000  & $2\times10^{9}$  & 0.03 \tabularnewline
$\mathit{b}$ \; \;  & 3000  & $200\times10^{9}$  & 0.04\tabularnewline
\hline 
\end{tabular}\vspace{0.3cm}
 \caption{\label{Table-1} Geometrical and physical properties of the phases
comprising the periodic beam. The width and thickness of the beam
are 0.02\,m and 0.0016\,m, respectively.}
\end{table}

\subsection{Homogenization of composite beams}

\begin{figure}[t!]
\centering \includegraphics[width=1.05\textwidth]{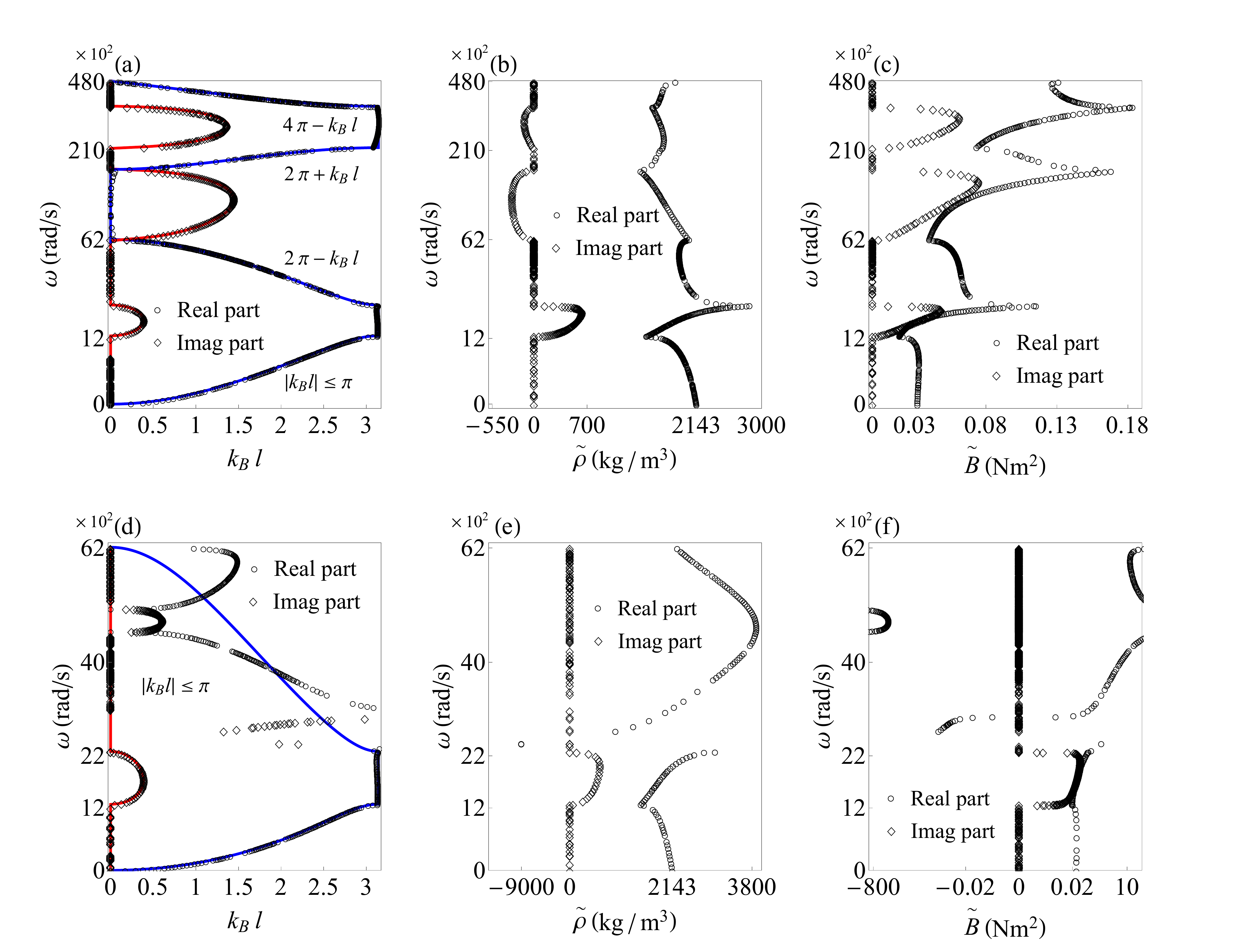}
\caption{ \label{Fig-2} (a) Band diagram of the exemplary composite, truncated
at the fifth band. The blue (resp. red) curve corresponds to the real
(resp. imaginary) part evaluated from the exact relation (\ref{Closed-E}).
Circle (resp. diamond) marks correspond to the real (resp. imaginary)
part evaluated from the homogenized dispersion relation (\ref{DRelation}).
(b) Effective dynamic mass density $\protect\rhoeff$ and (c) effective
dynamic bending stiffness $\protect\Beff$ employed in panel (a),
calculated using increasing values of $\protect\Blochwn$. Panels
(d-f) are the counterparts of (a-c), respectively, when the band diagram
is truncated at the second gap, and $\protect\Blochwn$ is restricted
to the $1^{\mathrm{st}}$ Brillouin zone. }
\end{figure}

\emph{Dispersion relation and effective parameters}. Consider an exemplary
periodic beam, whose phase geometrical and physical properties are
given in Tab. \ref{Table-1}. The first four bands, calculated by
the exact relation (\ref{Closed-E}), are given by the continuous
blue curves in Fig.\;\,\ref{Fig-2}(a). The imaginary part of $\Blochwn$,
associated with the gaps, is depicted by the continuous red curve.

Figs. \ref{Fig-2}(b) and \ref{Fig-2}(c) show the frequency-dependent
effective properties $\rhoeff$ and $\Beff$, respectively, calculated
using our scheme, \emph{i.e.}, via Eqs. (\ref{Beff-1}-\ref{Rhoeff}),
when truncating the Fourier series at $m=200$. Across the first band,
the effective properties are evaluated using $\Blochwn$ solutions
in the $1^{\mathrm{st}}$ Brillouin zone, while solutions in the $2^{\mathrm{nd}}$
(resp. $3^{\mathrm{rd}}$ and $4^{\mathrm{th}}$) Brillouin zone are
used across the second (resp. third and fourth) band range. We properly
obtain real values across the frequencies of the bands, and complex
values across the frequencies of the gaps. The homogenized dispersion
relation (\ref{DRelation}) is evaluated with these calculated effective
properties in Fig. \ref{Fig-2}(a), illustrated by the circle (real
part) and diamond (imaginary part) marks. Indeed, the exact and homogenized
dispersion relations are in excellent agreement. Notably, this agreement
extends beyond the fundamental Bloch band at low frequencies. 

As mentioned, if ($\omega,\Blochwn l$) satisfy the exact dispersion
relation, then so are ($\omega,2\pi n\pm\Blochwn l$) for integer
$n$. \citet{Willis2013} questioned if this ambiguity in the value
of $\Blochwn$ extends to homogenized models; we find that for the
homogenized dispersion relation with $\rhoeff$ and $\Beff$, only
a unique choice of $\Blochwn$ solution recovers the exact dispersion
relation. This is demonstrated in panels (d-f), where $\Blochwn$
values in the $1^{\mathrm{st}}$ Brillouin zone were chosen when calculating
$\rhoeff$, $\Beff$, and the dispersion relation across the second
band. Contrary to panels (a-c), were the choice of the $2^{\mathrm{nd}}$
Brillouin zone recovers the second band, here the homogenized dispersion
relation diverges from it. Similar divergence occurs at higher bands
where $\Blochwn$ solutions other than those employed in panel (d-f)
are used; for brevity, this illustration is omitted.

According to our formulation, the dynamic effective properties are
determined from Fourier coefficients $\rho(m)$, $B(m)$ which, for
a fixed unit cell length $l$, are independent of how the unit cell
of the periodic system is represented. Consequently, the curves reported
in Figs. \ref{Fig-2}(b) and \ref{Fig-2}(c) are also independent
of the unit cell representation. 
\begin{figure}[t!]
\centering \includegraphics[width=1.04\textwidth]{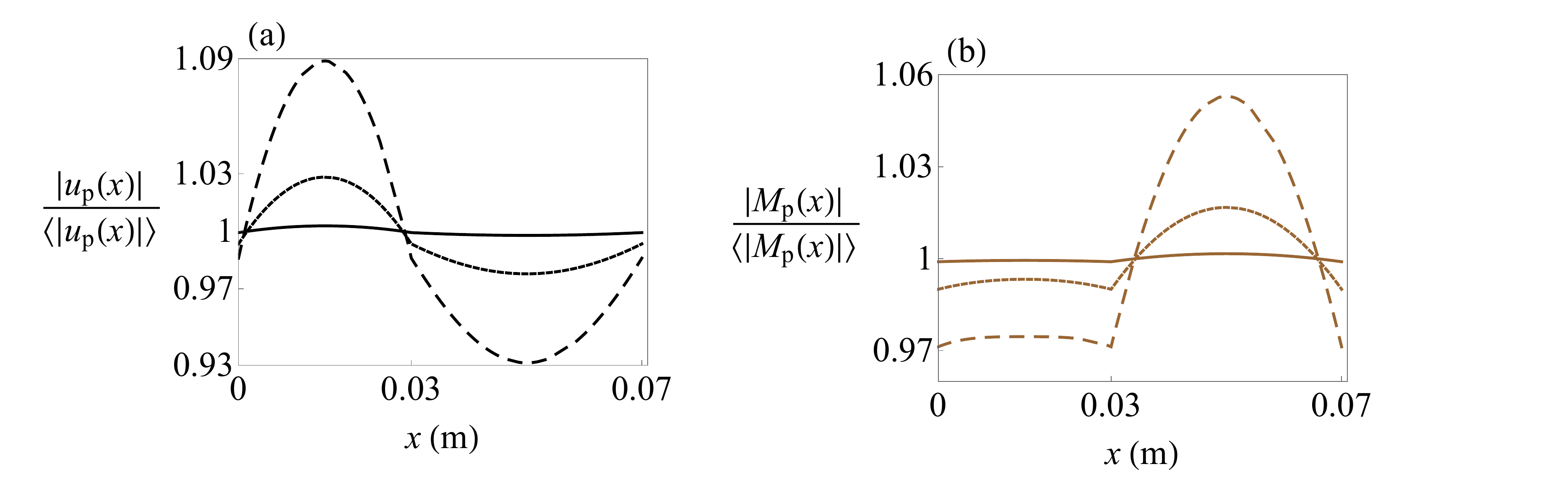}
\caption{ \label{Fig-3} The fields $\dfrac{|u_{\mathrm{p}}(x)|}{\langle|u_{\mathrm{p}}(x)|\rangle}$
(black curves), and $\dfrac{|M_{\mathrm{p}}(x)|}{\langle|M_{\mathrm{p}}(x)|\rangle}$
(brown curves) over the unit cell at 10, 100 and 300 $\dfrac{\mathrm{rad}}{\mathrm{s}}$
(solid, dotted and dashed curves, respectively).}
\end{figure}

\emph{Contiguous semi-infinite beams}. We continue to the interface
problem between the semi-infinite composite and its homogenized equivalent.
In Sec. \ref{Quantifying-applicability} we showed that the spatial
low variation of the displacement and bending moment fields at low
frequencies implies that $r\approx0$, and hence the homogenized model
captures the reflectance behavior of the periodic beam. As the frequency
is increased, the fields $u_{\mathrm{p}}(x)$ and $M_{\mathrm{p}}(x)$
fluctuate more rapidly, reflection becomes significant, and the suitability
of the homogenized model deteriorates. We demonstrate this in Fig.
\ref{Fig-3}, by plotting the normalized periodic part of the displacement
field (panel a), $\dfrac{|u_{\mathrm{p}}(x)|}{\langle|u_{\mathrm{p}}(x)|\rangle}$,
and the bending moment (panel b), $\dfrac{|M_{\mathrm{p}}(x)|}{\langle|M_{\mathrm{p}}(x)|\rangle}$,
over the unit cell, at the frequencies 10 $\dfrac{\mathrm{rad}}{\mathrm{s}}$
(solid curves), 100 $\dfrac{\mathrm{rad}}{\mathrm{s}}$ (dotted curves),
and 300 $\dfrac{\mathrm{rad}}{\mathrm{s}}$ (dashed curves). It can
be seen that the periodic parts of the displacement and the bending
moment are practically constants for frequencies below 100 $\dfrac{\mathrm{rad}}{\mathrm{s}}$,
while having a more significant variation at 300\,$\dfrac{\mathrm{rad}}{\mathrm{s}}$.
At this frequency, the normalized periodic part of the displacement
reach a maximum deviation of 0.09 from unity. In accordance with the
increase in mode fluctuation at higher frequencies, reflectance in
the semi-infinite interface problem increases too; for instance, when
considering an interface at the middle of the phase $a$, the reflection
at the frequencies 10, 100 and 300 $\dfrac{\mathrm{rad}}{\mathrm{s}}$
is $6.32\times10^{-7}$, $7.04\times10^{-5}$, and $7.84\times10^{-4}$,
respectively.

\begin{figure}[t!]
\centering \includegraphics[width=1.02\textwidth]{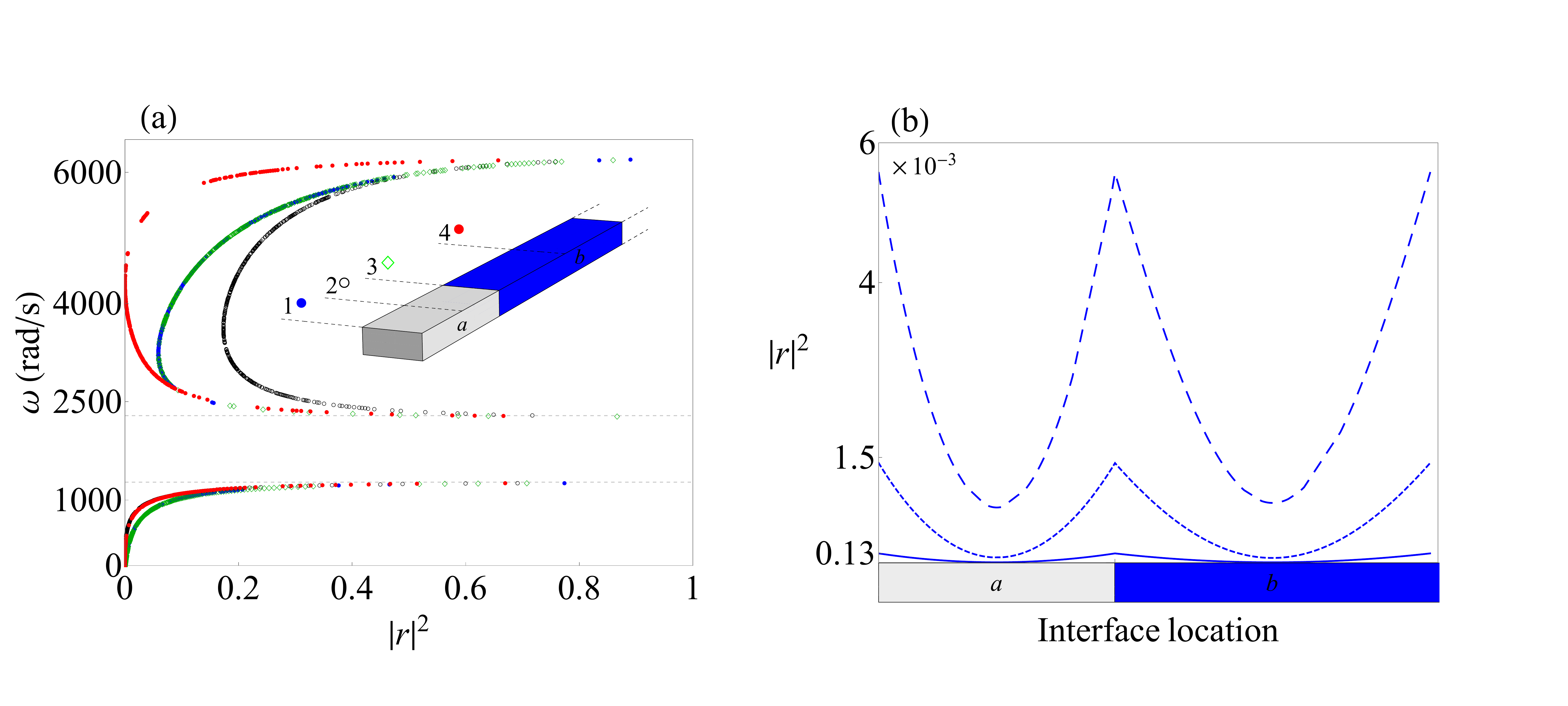}
\caption{ \label{Fig-4} (a) Frequency-reflected energy diagram for four locations
of the interface with the semi-infinite homogenized beam. (b) Reflected
energy as function of the interface location at 10, 100 and 300 $\dfrac{\mathrm{rad}}{\mathrm{s}}$
(solid, dotted and dashed blue curves, respectively). }
\end{figure}

The dependency of the reflected energy on the wavelength and interface
location is notably demonstrated in Fig.\;\ref{Fig-4}(a) by plotting
$\omega-\left|r\right|^{2}$ diagram for four different interface
locations, illustrated in the inset. Indeed, we observe that the reflected
energy vanishes when $\omega\to0$, and changes between different
interface locations. The difference is more pronounced across the
second band, specifically between the case of an interface at the
middle of phase $a$ and an interface at the middle of phase $b$.
Fig.\;\ref{Fig-4}(b) displays $\left|r\right|^{2}$ across the unit
cell at 10, 100 and 300 $\dfrac{\mathrm{rad}}{\mathrm{s}}$. We observe
that the reflection dependency on the interface location becomes greater
as the frequency increases. 

\begin{figure}[t!]
\centering \includegraphics[width=1.04\textwidth]{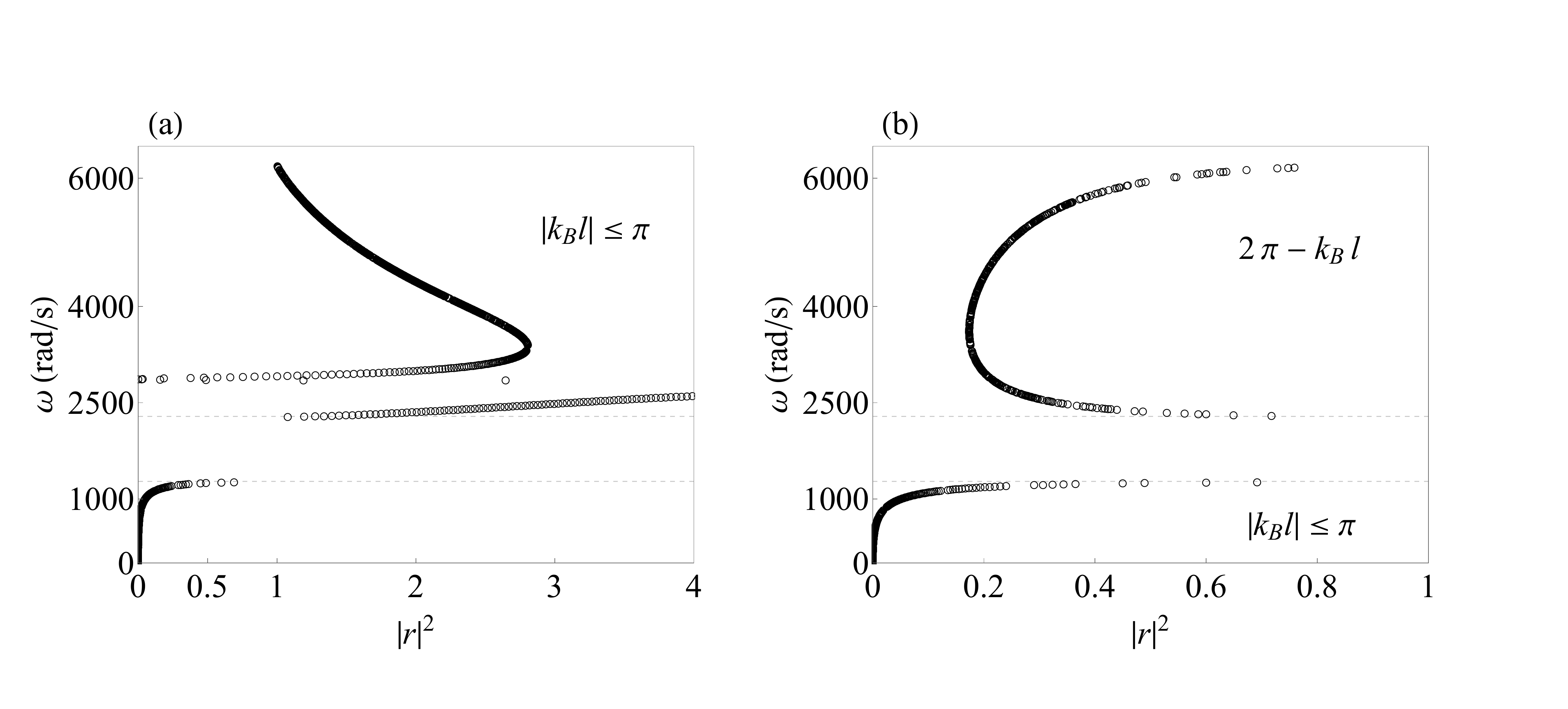}
\caption{ \label{Fig-5} Normalized reflected energy from the interface between
the exemplary semi-infinite periodic beam and its homogenized equivalent.
Panel (a) uses solutions in the $1^{\mathrm{st}}$ Brillouin zone
and for both branches. Panel (b) uses solutions in the $2^{\mathrm{nd}}$
Brillouin zone for the second branch.}
\end{figure}

We recall that in calculating $\rhoeff$ and $\Beff$, and in turn
$r$, the right Brillouin zone should be chosen. To demonstrate it
in this problem, we plot in Fig.\;\,\ref{Fig-5} the normalized
reflected energy of the exemplary beam, when it is in contact with
its homogenized equivalent at $x=0$. We evaluate $\left|r\right|^{2}$
across the frequency range of the first two bands, when the interface
is located at the middle of phase $a$. In Fig.\;\,\ref{Fig-5}(a),
the calculation was carried out using solutions in the $1^{\mathrm{st}}$
Brillouin zone, while Fig.\;\,\ref{Fig-5}(b) uses solutions in
the $2^{\mathrm{nd}}$ Brillouin zone for the second band range. The
reflected energy is found to be independent of whether we choose the
positive normalized wavenumber or its negative value. We observe that
calculating $r$ across the second band using $\Blochwn l$ in the
$1^{\mathrm{st}}$ Brillouin zone violets the conservation of energy,
\textit{i.e}., $\left|r\right|^{2}>1$. By contrast, choosing solutions
in the $2^{\mathrm{nd}}$ Brillouin zone leads to $\left|r\right|^{2}<1$,
as it physically should. These results agree with our discussion following
Fig. \ref{Fig-2}, as well as \citet{Srivastava2014} observation,
that certain homogenization schemes for Bloch waves must use proper
Brillouin zones. 

\begin{figure}[t!]
\centering \includegraphics[width=1.04\textwidth]{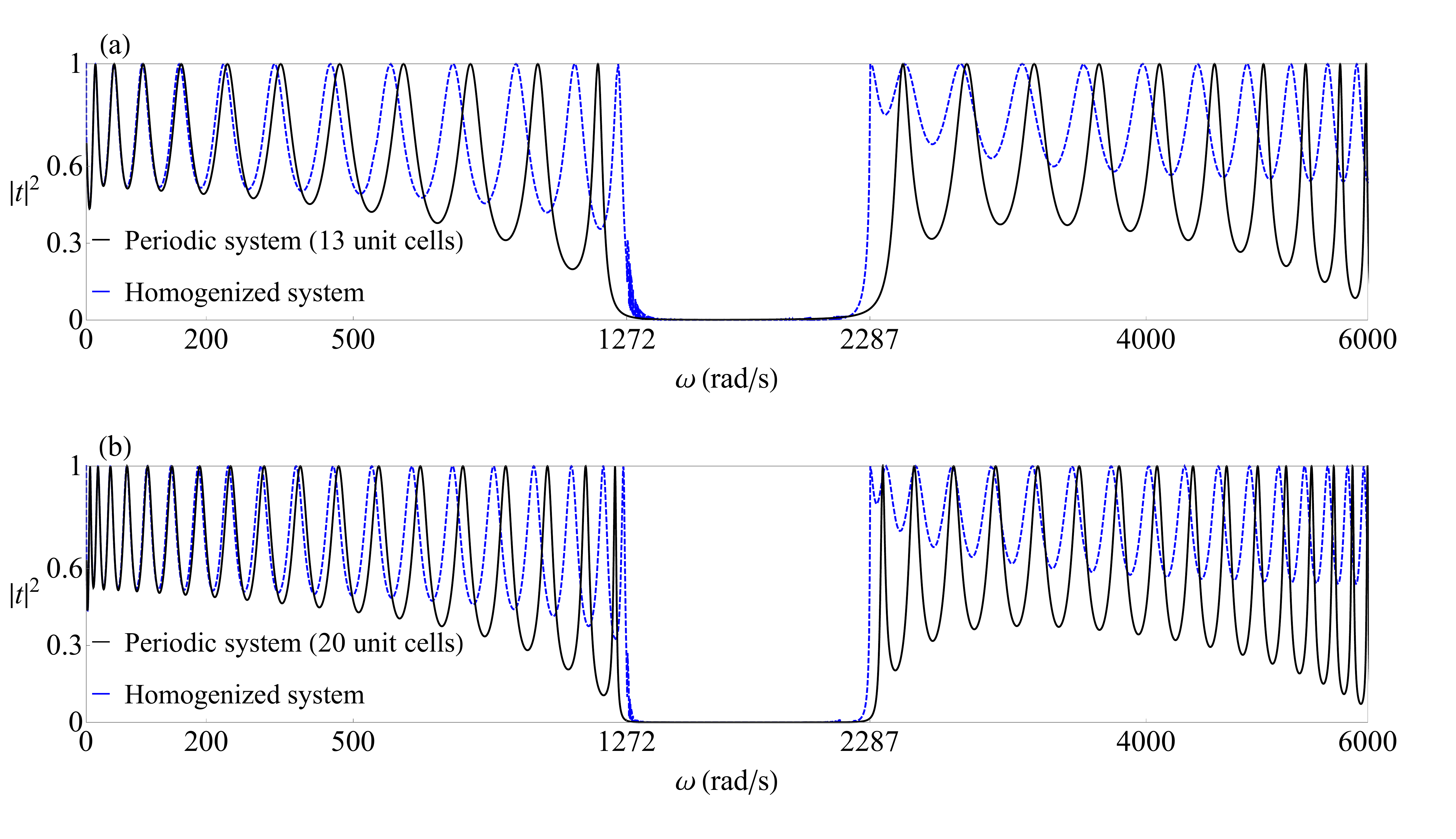}
\caption{ \label{Fig-6} Normalized transmitted energy through two semi-infinite
$b$ phases connected by a finite periodic beam (solid black curves),
and when connected by the homogenized beam of the same length (dashed
blue curves). The length of the intermediate beam in panels (a) and
(b) is of 13 and 20 unit cells, respectively.}
\end{figure}

\emph{Finite beam bounded by two semi-infinite homogeneous beams}.
Next, we consider a composite comprising a finite number of unit cells,
and compare its transmission spectrum with the spectrum of a homogeneous
beam of the same length, whose properties are the composite homogenized
properties, see Fig.\,\ref{Fig-1}(d). Fig.\;\,\ref{Fig-6} shows
the normalized transmitted energy through two semi-infinite $b$ phases
when they are connected by a finite periodic beam (solid black curves),
and when connected by the homogenized beam of the same length (dashed
blue curves). In Fig. \ref{Fig-6}(a), the calculation was carried
out for an intermediate beam comprising 13 unit cells, while Fig.
\ref{Fig-6}(b) depicts the result for an intermediate beam comprising
20 unit cells. At low frequencies, the homogenized model reproduces
almost identically the transmission characteristics of the periodic
beam. For instance, the peak frequencies differ in less than 2\% in
the range $0-200$ $\dfrac{\mathrm{rad}}{\mathrm{s}}$ for the shorter
beams. For longer beams, the difference is even smaller; across the
same frequency range, the minimal values of the transmitted energy
differ in less than 1.3\%. Across the frequency range of the second
band, the homogenized model still reasonably recovers peak locations.
However, the error in predicting the minimal transmitted energy becomes
substantial, \emph{e.g.}, in the second band depicted in panel (b),
the highest value of this error is 87\%. 

\subsection{Homogenization of locally resonant beams}

Next, we apply our scheme to a uniform aluminum beam with periodically
attached local resonators. The beam properties are $\rho=2700$\,$\dfrac{\mathrm{kg}}{\mathrm{m}^{3}}$,
$E=70\times10^{9}$\,$\mathrm{Pa}$ and the cross-section of the
beam (width$\times$ thickness) is $0.03\times0.025$\,$\mathrm{m^{2}}$.
The local resonators properties are $k_{s}=1.455\times10^{6}$\,$\dfrac{\mathrm{N}}{\mathrm{m}}$,
$m=0.069$\,$\mathrm{kg}$, as in \citet{2013-Xiao-periodic-beams-vibration-absorbers};
the distance between the resonators is $l=0.04$\,$\mathrm{m}$.
These values correspond to a resonance frequency of $\omega_{0}=4582.8$\,$\dfrac{\mathrm{rad}}{\mathrm{s}}$.

\begin{figure}[t!]
\centering \includegraphics[width=1.03\textwidth]{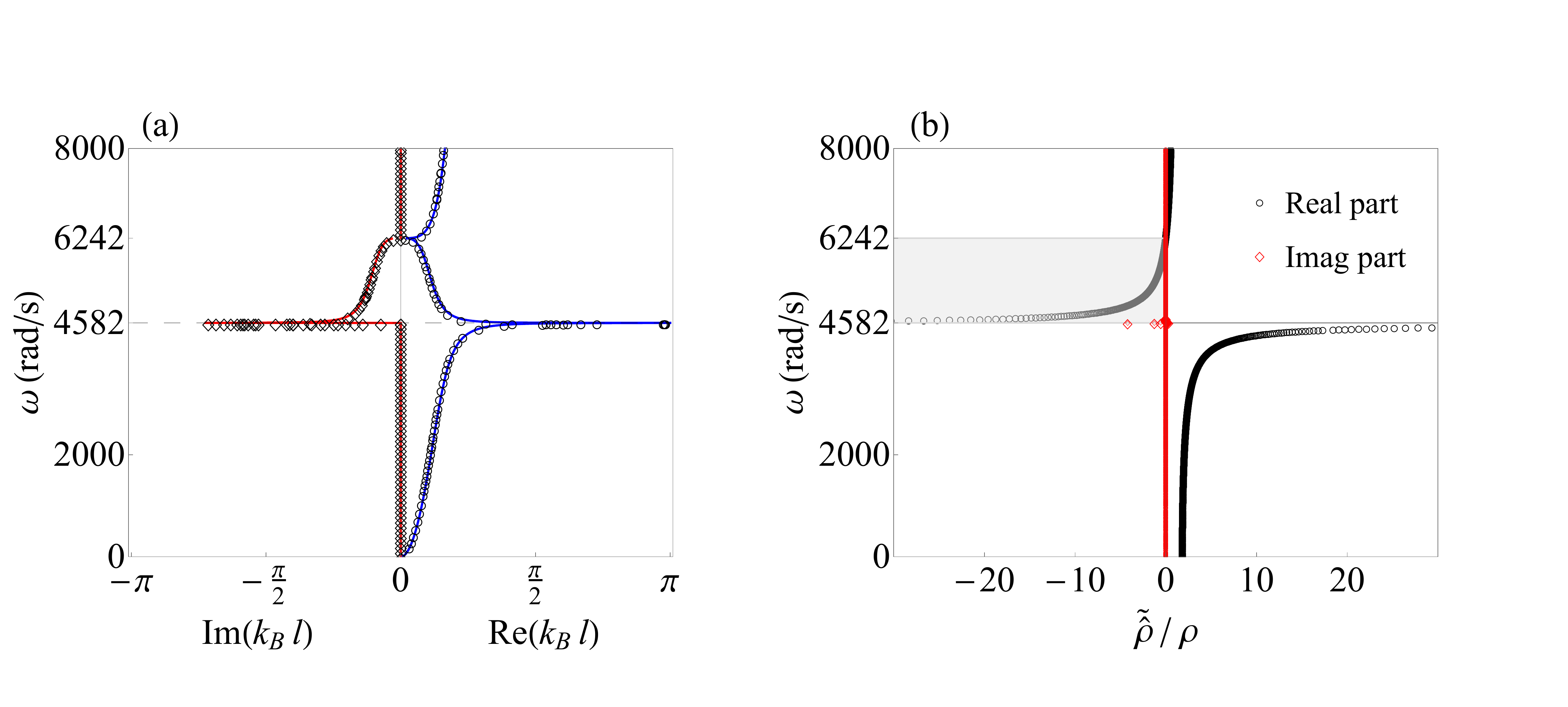}
\caption{ \label{Fig-7} (a) Band structure of the uniform beam with local
resonators. The blue (resp. red) curve corresponds to the real (resp.
imaginary) part evaluated from the exact relation (\ref{Closed-E}).
Circle (resp. diamond) marks correspond to the real (resp. imaginary)
part evaluated from the homogenized dispersion relation (\ref{DRelation}).
(b) Normalized effective mass density.}
\end{figure}

\emph{Dispersion relation and effective properties}. Firstly, we evaluate
in Fig.\;\ref{Fig-7}(a) the exact dispersion relation (continuous
curves) using Eq.\;(\ref{Closed-E}). Subsequently, we calculate
$\tilde{\hat{\rho}}$ and the homogenized dispersion relation using
Eqs.\;(\ref{DRelation}) and (\ref{Rhoeff}), when the Fourier expansion
comprises 40 terms. The effective mass density, normalized by $\rho$,
is depicted in Fig.\;\ref{Fig-7}(b), where black circle marks correspond
to its real part and red diamond marks correspond to its imaginary
part. Note that in this case we used $\left|\Blochwn l\right|\leq\pi$;
the corresponding homogenized dispersion relation is depicted by circle
(real part) and diamond (imaginary part) marks in Fig.\;\ref{Fig-7}(a),
and demonstrates an excellent agreement with the exact relation. Both
relations exhibit a locally resonant gap across the range ${4582.8\,\dfrac{\mathrm{rad}}{\mathrm{s}}<\omega<6242.2\,\dfrac{\mathrm{rad}}{\mathrm{s}}}$.
Note that the frequency at which the gap opens is independent of $l$,
a known feature of locally resonant gaps. 

\begin{figure}[t!]
\centering \includegraphics[width=1.04\textwidth]{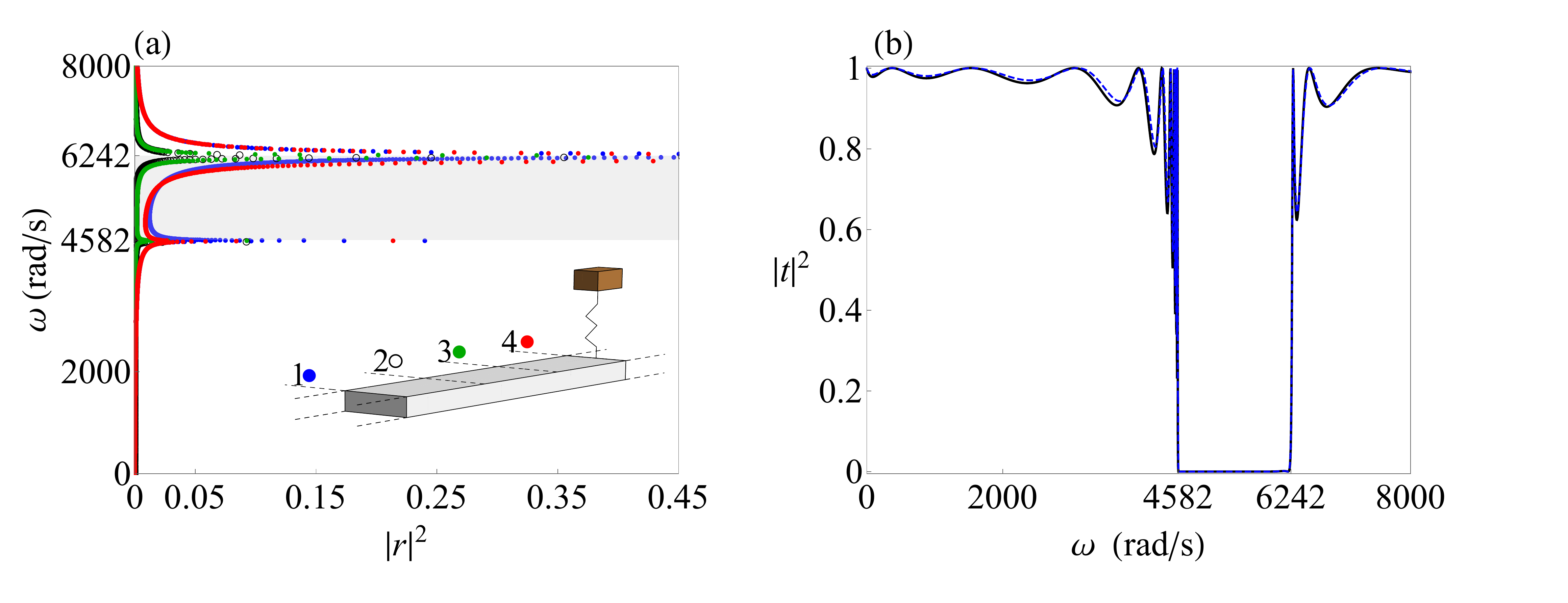}
\caption{\label{Fig-8} (a) Frequency-reflected energy diagram for four equi-spaced
locations of the interface with the semi-infinite homogenized beam.
The dashed area indicates the metamaterial region, in which the effective
mass density is negative. (b) Normalized transmitted energy through
two semi-infinite beams made of aluminum, connected by a locally resonant
beam comprising 20 unit cells (solid black curve), and when connected
by a homogenized beam of the same length (dashed blue curve).}
\end{figure}

\emph{Contiguous semi-infinite beams}. Fig.\;\,\ref{Fig-8}(a) shows
the normalized reflected energy of the interface problem between the
exemplary semi-infinite beam with local resonators and its semi-infinite
homogenized beam for four equi-spaced interface locations, illustrated
in the inset. We observe a dependency of the reflected energy with
the interface location which is greater in the vicinity of the frequencies
at which $\tilde{\hat{\rho}}=0$ and $\infty$. At these frequencies,
the difference in the reflected energy is most pronounced between
the case of an interface located at one of the ends of the unit cell
and an interface near its center. We observe that the homogenized
model of the resonant system is less sensitive to the interface location
than the model for the composite beam. We further observe that except
in the vicinity of the frequencies at which $\tilde{\hat{\rho}}=0$
and $\infty$, the normalized reflected energy vanishes, indicating
that the applied homogenization scheme is a valid approximation across
a wide range of frequencies. This observation\textemdash that locally
resonant media are approximated better than non-resonant media by
homogenization\textemdash was noted before, \emph{e.g.}, by \citet{Srivastava2014},
and \citet{2015-Elastic-metamaterials-and-dynamic-homogenization-review}.

\emph{Finite beam bounded by two semi-infinite homogeneous beams}.
Fig.\;\,\ref{Fig-8}(b) shows the normalized transmitted energy
through two semi-infinite beams made of aluminum, connected by a locally
resonant beam comprising 20 unit cells (solid black curve), and when
connected by a homogenized beam of the same length (dashed blue curve).
Remarkably, the homogenized model reproduces almost identically the
transmission of the periodic beam. Here again, we find that the homogenization
of the locally resonant medium reproduces the periodic medium characteristics
better the characteristics of the the composite medium. 

\section{\label{Conclusion}Summary}

We have developed a new homogenized model for composite beams and
systems with periodically attached local resonators undergoing flexural
motions, based on the approach of \citet{NematNasser2011jmps}. Specifically,
we derived macroscopic equations---which take the same form of the
local ones---and effective properties, which produce together the
exact dispersion relation, as desired. In addition to the simplicity
of our homogenization scheme, it does not require the knowledge of
local fields; these are actually extractable from it, if wanted. 

We have investigated the capability of the homogenized model to capture
the dynamic characteristics of the periodic systems, through its application
in three numerical settings. Firstly, we have compared the exact dispersion
relation of infinite exemplary composite and locally resonant systems
with our homogenized model, to find an excellent agreement. Secondly,
we studied the reflected energy of an incident wave from a semi-infinite
homogenized medium to its semi-infinite periodic counterpart. We showed
that at low frequencies, there is no reflection from the media common
interface, namely, in this limit our model is able to appropriately
match the periodic system impedance. As frequency increases, impedances
cannot be matched, and reflection occur in a manner that depends on
microscale details, namely, the impedance and length of the phase
that borders with the homogenized medium. We observed that this dependency
is weaker in the locally resonant case. Finally, we have analyzed
the transmitted energy of an incident wave through a finite medium
bounded between two semi-infinite homogeneous beams. Specifically,
we have compared the transmitted energy when the intermediate medium
is periodic, with the transmitted energy when the periodic medium
it is replaced by its homogenized equivalent. We observed that the
homogenized model of the composite beam neatly captures the first
gap, as well as the transmission spectrum across the first band, with
some deterioration towards its end. Across the second band, the homogenized
model predicts well frequencies of complete transmission, while substantially
overestimating magnitudes of minimal transmission. The homogenized
approximation for the locally resonant beam is significantly better,
having its spectrum almost indistinguishable from the periodic medium
spectrum across the two bands. This observation, together with a similar
observation in the case of semi-infinite media, suggests that locally
resonant systems lend themselves to homogenization better than composite
systems, as noted by \citet{Srivastava2014}. It is imperative to
extend our analysis to other structural models, such as plates \citep{2012-High-frequency-plates-FWaves}
and torsional systems \citep{carta15brun}; this will be pursued in
future work.

\section*{Acknowledgements }

We acknowledge the supports of the Israel Science Foundation, funded
by the Israel Academy of Sciences and Humanities (Grant no. 1912/15),
and the United States-Israel Binational Science Foundation (Grant
no. 2014358).

\appendix

\section*{Appendix A}

\label{Appendix-A} \global\long\def\theequation{A.\arabic{equation}}
 \setcounter{equation}{0}

Within a homogeneous phase, the state vector $\mathsf{s}(x)=\left\lbrace u(x),\theta(x),M(x),V(x)\right\rbrace ^{\mathsf{T}}$
can be expressed in the matrix form 
\begin{eqnarray}
\mathsf{s}(x) & = & \left[\begin{array}{cccc}
e^{i\waven x} & e^{-\waven x} & e^{-i\waven x} & e^{\waven x}\\
i\waven e^{i\waven x} & -\waven e^{-\waven x} & -i\waven e^{-i\waven x} & \waven e^{\waven x}\\
-B\waven^{2}e^{i\waven x} & B\waven^{2}e^{-\waven x} & -B\waven^{2}e^{-i\waven x} & B\waven^{2}e^{\waven x}\\
iB\waven^{3}e^{i\waven x} & B\waven^{3}e^{-\waven x} & -iB\waven^{3}e^{-i\waven x} & -B\waven^{3}e^{\waven x}
\end{array}\right]\cdot\left\lbrace \begin{array}{c}
C^{+}\\
D^{+}\\
C^{-}\\
D^{-}
\end{array}\right\rbrace ,\label{s-vector}
\end{eqnarray}
where $\waven=\sqrt[4]{\rho A\omega^{2}/B}$ and the coefficients
$C^{\pm},D^{\pm}$ represent amplitudes of corresponding waves. Commonly,
the state vector in periodic systems, and in turn the dispersion relation,
are expressed in terms of the transfer matrix. However, in certain
cases the transfer matrix formulation is prone to numerical instabilities
\citep{Dunkin1965,Perez2004}. Therefore, we used a formulation based
on the numerically stable hybrid matrix \citep{tan06hybrid,rene15siam,2016-Shmuel-SMS},
as follow. We define a modified state vector $\mathsf{s}_{\mathrm{m}}(x)=\left\lbrace u(x),\theta(x)\waven^{(a)^{-1}},M(x)\waven^{(a)^{-2}},V(x)\waven^{(a)^{-3}}\right\rbrace ^{\mathsf{T}}$,
for which the hybrid matrix reads in Eq. (\ref{definition-Hyb-matrix})
{\small{}
\begin{eqnarray}
\mathsf{H}^{(i)} & = & \left[\begin{array}{cccc}
1 & 1 & e^{i\waven^{(i)}\dl} & e^{-\waven^{(i)}\dl}\\[3pt]
i\dfrac{\waven^{(i)}}{\waven^{(a)}} & -\dfrac{\waven^{(i)}}{\waven^{(a)}} & -i\dfrac{\waven^{(i)}}{\waven^{(a)}}e^{i\waven^{(i)}\dl} & \dfrac{\waven^{(i)}}{\waven^{(a)}}e^{-\waven^{(i)}\dl}\\[10pt]
-B^{(i)}\left(\dfrac{\waven^{(i)}}{\waven^{(a)}}\right)^{2}e^{i\waven^{(i)}\dl} & B^{(i)}\left(\dfrac{\waven^{(i)}}{\waven^{(a)}}\right)^{2}e^{-\waven^{(i)}\dl} & -B^{(i)}\left(\dfrac{\waven^{(i)}}{\waven^{(a)}}\right)^{2} & B^{(i)}\left(\dfrac{\waven^{(i)}}{\waven^{(a)}}\right)^{2}\\[15pt]
iB^{(i)}\left(\dfrac{\waven^{(i)}}{\waven^{(a)}}\right)^{3}e^{i\waven^{(i)}\dl} & B^{(i)}\left(\dfrac{\waven^{(i)}}{\waven^{(a)}}\right)^{3}e^{-\waven^{(i)}\dl} & -iB^{(i)}\left(\dfrac{\waven^{(i)}}{\waven^{(a)}}\right)^{3} & -B^{(i)}\left(\dfrac{\waven^{(i)}}{\waven^{(a)}}\right)^{3}
\end{array}\right]\nonumber \\
 &  & \cdot\left[\begin{array}{cccc}
-B^{(i)}\left(\dfrac{\waven^{(i)}}{\waven^{(a)}}\right)^{2} & B^{(i)}\left(\dfrac{\waven^{(i)}}{\waven^{(a)}}\right)^{2} & -B^{(i)}\left(\dfrac{\waven^{(i)}}{\waven^{(a)}}\right)^{2}e^{i\waven^{(i)}\dl} & B^{(i)}\left(\dfrac{\waven^{(i)}}{\waven^{(a)}}\right)^{2}e^{-\waven^{(i)}\dl}\\[15pt]
iB^{(i)}\left(\dfrac{\waven^{(i)}}{\waven^{(a)}}\right)^{3} & B^{(i)}\left(\dfrac{\waven^{(i)}}{\waven^{(a)}}\right)^{3} & -iB^{(i)}\left(\dfrac{\waven^{(i)}}{\waven^{(a)}}\right)^{3}e^{i\waven^{(i)}\dl} & -B^{(i)}\left(\dfrac{\waven^{(i)}}{\waven^{(a)}}\right)^{3}e^{-\waven^{(i)}\dl}\\[15pt]
e^{i\waven^{(i)}\dl} & e^{-\waven^{(i)}\dl} & 1 & 1\\[3pt]
i\dfrac{\waven^{(i)}}{\waven^{(a)}}e^{i\waven^{(i)}\dl} & -\dfrac{\waven^{(i)}}{\waven^{(a)}}e^{-\waven^{(i)}\dl} & -i\dfrac{\waven^{(i)}}{\waven^{(a)}} & \dfrac{\waven^{(i)}}{\waven^{(a)}}
\end{array}\right]^{-1},\nonumber \\
\end{eqnarray}
}{\small \par}

\noindent where $B^{(i)}$ denotes the bending stiffness of the phase
$i$, and the $\waven^{(i)}$ (resp. $\waven^{(a)}$) is the value
that $\waven$ takes in the phase $i$ (resp. $a$). The matrix elements
of $\mathsf{H}^{(i)}$ are 
\begin{eqnarray}
\begin{aligned}\mathsf{h}_{11}^{(i)} & =\left(\dfrac{\waven^{(a)}}{\waven^{(i)}}\right)^{2}\frac{\sin\,\dl\waven^{(i)}\sinh\,\dl\waven^{(i)}}{B^{(i)}\cos\,\dl\waven^{(i)}\cosh\,\dl\waven^{(i)}+B^{(i)}},\\[4pt]
\mathsf{h}_{12}^{(i)} & =\left(\dfrac{\waven^{(a)}}{\waven^{(i)}}\right)^{3}\frac{\cos\,\dl\waven^{(i)}\sinh\,\dl\waven^{(i)}-\sin\,\dl\waven^{(i)}\cosh\,\dl\waven^{(i)}}{B^{(i)}\cos\,\dl\waven^{(i)}\cosh\,\dl\waven^{(i)}+B^{(i)}},\\[4pt]
\mathsf{h}_{13}^{(i)} & =\dfrac{\cos\,\dl\waven^{(i)}+\cosh\,\dl\waven^{(i)}}{\cos\,\dl\waven^{(i)}\cosh\,\dl\waven^{(i)}+1},\\[4pt]
\mathsf{h}_{14}^{(i)} & =-\dfrac{\waven^{(a)}}{\waven^{(i)}}\frac{\sin\,\dl\waven^{(i)}+\sinh\,\dl\waven^{(i)}}{\cos\,\dl\waven^{(i)}\cosh\,\dl\waven^{(i)}+1},\\[15pt]
\end{aligned}
\end{eqnarray}
\begin{eqnarray}
\begin{aligned}\mathsf{h}_{21}^{(i)} & =-\dfrac{\waven^{(a)}}{\waven^{(i)}}\frac{\cos\dl\waven^{(i)}\sinh\dl\waven^{(i)}+\sin\dl\waven^{(i)}\cosh\dl\waven^{(i)}}{B^{(i)}\cos\dl\waven^{(i)}\cosh\dl\waven^{(i)}+B^{(i)}},\\[4pt]
\mathsf{h}_{22}^{(i)} & =\mathsf{h}_{11}^{(i)},\\[4pt]
\mathsf{h}_{23}^{(i)} & =\dfrac{\waven^{(i)}}{\waven^{(a)}}\frac{\sin\,\dl\waven^{(i)}-\sinh\,\dl\waven^{(i)}}{\cos\,\dl\waven^{(i)}\cosh\,\dl\waven^{(i)}+1},\\[4pt]
\mathsf{h}_{24}^{(i)} & =\mathsf{h}_{13}^{(i)},\\[15pt]
\end{aligned}
\end{eqnarray}
\vspace{0.05cm}
 
\begin{eqnarray}
\begin{aligned}\mathsf{h}_{31}^{(i)} & =\mathsf{h}_{13}^{(i)},\\[4pt]
\mathsf{h}_{32}^{(i)} & =\mathsf{h}_{14}^{(i)},\\[4pt]
\mathsf{h}_{33}^{(i)} & =\left(\dfrac{\waven^{(i)}}{\waven^{(a)}}\right)^{2}\frac{B^{(i)}\sin\,\dl\waven^{(i)}\sinh\,\dl\waven^{(i)}}{\cos\,\dl\waven^{(i)}\cosh\,\dl\waven^{(i)}+1},\\[4pt]
\mathsf{h}_{34}^{(i)} & =\dfrac{\waven^{(i)}}{\waven^{(a)}}\frac{B^{(i)}[\cos\,\dl\waven^{(i)}\sinh\,\dl\waven^{(i)}-\sin\,\dl\waven^{(i)}\cosh\,\dl\waven^{(i)}]}{\cos\,\dl\waven^{(i)}\cosh\,\dl\waven^{(i)}+1},\\[15pt]
\end{aligned}
\end{eqnarray}
\vspace{0.05cm}
 
\begin{eqnarray}
\begin{aligned}\mathsf{h}_{41}^{(i)} & =\mathsf{h}_{23}^{(i)},\\[4pt]
\mathsf{h}_{42}^{(i)} & =\mathsf{h}_{13}^{(i)},\\[4pt]
\mathsf{h}_{43}^{(i)} & =-\left(\dfrac{\waven^{(i)}}{\waven^{(a)}}\right)^{3}\frac{B^{(i)}[\cos\,\dl\waven^{(i)}\sinh\,\dl\waven^{(i)}+\sin\dl\waven^{(i)}\cosh\,\dl\waven^{(i)}]}{\cos\,\dl\waven^{(i)}\cosh\,\dl\waven^{(i)}+1},\\[4pt]
\mathsf{h}_{44}^{(i)} & =\mathsf{h}_{33}^{(i)}.
\end{aligned}
\end{eqnarray}
The total hybrid matrix corresponding to a stack comprising $i$ phases,
$\bH^{(1,i)}$, is determined in terms of the total hybrid matrix
of the first $i-1$ phases, $\bH^{(1,i-1)}$, and the hybrid matrix
of the $i$th phase, $\bH^{(i)}$, as follows 
\begin{equation}
\begin{alignedat}{1}\bH_{22}^{(1,i)} & =\bH_{22}^{(i)}+\bH_{21}^{(i)}\cdot\left[\bI-\bH_{22}^{(1,i-1)}\cdot\bH_{11}^{\left(i\right)}\right]^{-1}\cdot\bH_{22}^{(1,i-1)}\cdot\bH_{12}^{\left(i\right)},\\
\bH_{21}^{(1,i)} & =\bH_{21}^{\left(i\right)}\cdot\left[\bI-\bH_{22}^{(1,i-1)}\cdot\bH_{11}^{\left(i\right)}\right]^{-1}\cdot\bH_{21}^{\left(1,i-1\right)},\\
\bH_{12}^{(1,i)} & =\bH_{12}^{(1,i-1)}\cdot\bH_{12}^{\left(i\right)}+\bH_{12}^{(1,i-1)}\cdot\bH_{11}^{\left(i\right)}\cdot\left[\bI-\bH_{22}^{(1,i-1)}\cdot\bH_{11}^{\left(i\right)}\right]^{-1}\cdot\bH_{22}^{(1,i-1)}\cdot\bH_{12}^{\left(i\right)},\\
\bH_{11}^{(1,i)} & =\bH_{11}^{(1,i-1)}+\bH_{12}^{(1,i-1)}\cdot\bH_{11}^{\left(i\right)}\cdot\left[\bI-\bH_{22}^{(1,i-1)}\cdot\bH_{11}^{\left(i\right)}\right]^{-1}\cdot\bH_{21}^{(1,i-1)},
\end{alignedat}
\label{eq: composition rule for H}
\end{equation}
where $\mathsf{H}_{11},\mathsf{H}_{12},\mathsf{H}_{21},\mathsf{H}_{22}$
denote the $2\times2$ sub-blocks of the corresponding hybrid matrix,
and $\bI$ denotes the unit matrix of the same order.

\noindent 
 \bibliographystyle{plainnat}


\begin{thebibliography}{44}
\providecommand{\natexlab}[1]{#1}
\providecommand{\url}[1]{\texttt{#1}}
\expandafter\ifx\csname urlstyle\endcsname\relax
  \providecommand{\doi}[1]{doi: #1}\else
  \providecommand{\doi}{doi: \begingroup \urlstyle{rm}\Url}\fi

\bibitem[Amirkhizi(2017)]{AMIRKHIZI2017}
Alireza~V. Amirkhizi.
\newblock Homogenization of layered media based on scattering response and
  field integration.
\newblock \emph{Mechanics of Materials}, 114:\penalty0 76 -- 87, 2017.
\newblock ISSN 0167-6636.
\newblock \doi{https://doi.org/10.1016/j.mechmat.2017.06.008}.
\newblock URL
  \url{http://www.sciencedirect.com/science/article/pii/S0167663617302016}.

\bibitem[Antonakakis and Craster(2012)]{2012-High-frequency-plates-FWaves}
T.~Antonakakis and R.~V. Craster.
\newblock High-frequency asymptotics for microstructured thin elastic plates
  and platonics.
\newblock \emph{Proceedings of the Royal Society of London A: Mathematical,
  Physical and Engineering Sciences}, 468\penalty0 (2141):\penalty0 1408--1427,
  2012.
\newblock ISSN 1364-5021.
\newblock \doi{10.1098/rspa.2011.0652}.
\newblock URL
  \url{http://rspa.royalsocietypublishing.org/content/468/2141/1408}.

\bibitem[Barnwell et~al.(2017)Barnwell, Parnell, and
  Abrahams]{Barnwell2017parnell}
E.~G. Barnwell, W.~J. Parnell, and I.~D. Abrahams.
\newblock Tunable elastodynamic band gaps.
\newblock \emph{Extreme Mechanics Letters}, 12:\penalty0 23--29, 2017.
\newblock ISSN 2352-4316.
\newblock \doi{http://doi.org/10.1016/j.eml.2016.10.009}.
\newblock URL
  \url{http://www.sciencedirect.com/science/article/pii/S2352431616300815}.
\newblock Frontiers in Mechanical Metamaterials.

\bibitem[Bigoni et~al.(2013)Bigoni, Guenneau, Movchan, and Brun]{bigoni2013prb}
D.~Bigoni, S.~Guenneau, A.~B. Movchan, and M.~Brun.
\newblock Elastic metamaterials with inertial locally resonant structures:
  Application to lensing and localization.
\newblock \emph{Phys. Rev. B}, 87:\penalty0 174303, May 2013.
\newblock \doi{10.1103/PhysRevB.87.174303}.
\newblock URL \url{https://link.aps.org/doi/10.1103/PhysRevB.87.174303}.

\bibitem[Bloch(1929)]{1928-Bloch-Zeit-fur-Physik}
F.~Bloch.
\newblock \emph{Z. Physik}, 52:\penalty0 555, 1929.
\newblock \doi{https://doi.org/10.1007/BF01339455}.

\bibitem[Carta and Brun(2015)]{carta15brun}
G.~Carta and M.~Brun.
\newblock Bloch--floquet waves in flexural systems with continuous and discrete
  elements.
\newblock \emph{Mechanics of Materials}, 87:\penalty0 11--26, 2015.
\newblock ISSN 0167-6636.
\newblock \doi{http://dx.doi.org/10.1016/j.mechmat.2015.03.004}.
\newblock URL
  \url{http://www.sciencedirect.com/science/article/pii/S0167663615000678}.

\bibitem[Celli and Gonella(2014)]{celli14}
P.~Celli and S.~Gonella.
\newblock Low-frequency spatial wave manipulation via phononic crystals with
  relaxed cell symmetry.
\newblock \emph{Journal of Applied Physics}, 115\penalty0 (10):\penalty0
  103502, 2014.
\newblock \doi{10.1063/1.4867918}.
\newblock URL \url{https://doi.org/10.1063/1.4867918}.

\bibitem[Cerd\'an-Ram\'{\i}rez et~al.(2009)Cerd\'an-Ram\'{\i}rez,
  Zenteno-Mateo, Sampedro, Palomino-Ovando, Flores-Desirena, and
  P\'erez-Rodr\'{\i}guez]{2009-Felipe-GetPDFServlet}
V.~Cerd\'an-Ram\'{\i}rez, B.~Zenteno-Mateo, M.~P. Sampedro, M.~A.
  Palomino-Ovando, B.~Flores-Desirena, and F.~P\'erez-Rodr\'{\i}guez.
\newblock Anisotropy effects in homogenized magnetodielectric photonic
  crystals.
\newblock \emph{Journal of Applied Physics}, 106\penalty0 (10):\penalty0
  103520, 2009.
\newblock \doi{10.1063/1.3261758}.
\newblock URL \url{https://doi.org/10.1063/1.3261758}.

\bibitem[Chen et~al.(2017)Chen, Hu, and Huang]{CHEN2017179}
Y.~Chen, G.~Hu, and G.~Huang.
\newblock A hybrid elastic metamaterial with negative mass density and tunable
  bending stiffness.
\newblock \emph{Journal of the Mechanics and Physics of Solids}, 105:\penalty0
  179 -- 198, 2017.
\newblock ISSN 0022-5096.
\newblock \doi{https://doi.org/10.1016/j.jmps.2017.05.009}.
\newblock URL
  \url{http://www.sciencedirect.com/science/article/pii/S0022509617301229}.

\bibitem[Colquitt et~al.(2014)Colquitt, Brun, Gei, Movchan, Movchan, and
  Jones]{Colquitt2014}
D.J. Colquitt, M.~Brun, M.~Gei, A.B. Movchan, N.V. Movchan, and I.S. Jones.
\newblock Transformation elastodynamics and cloaking for flexural waves.
\newblock \emph{Journal of the Mechanics and Physics of Solids}, 72:\penalty0
  131--143, 2014.
\newblock ISSN 0022-5096.
\newblock \doi{http://dx.doi.org/10.1016/j.jmps.2014.07.014}.
\newblock URL
  \url{http://www.sciencedirect.com/science/article/pii/S0022509614001586}.

\bibitem[Craster et~al.(2010)Craster, Kaplunov, and
  Pichugin]{2010-Craster-PRSA}
R.~V. Craster, J.~Kaplunov, and A.~V. Pichugin.
\newblock High-frequency homogenization for periodic media.
\newblock \emph{Proceedings of the Royal Society of London A: Mathematical,
  Physical and Engineering Sciences}, 466\penalty0 (2120):\penalty0 2341--2362,
  2010.
\newblock ISSN 1364-5021.
\newblock \doi{10.1098/rspa.2009.0612}.
\newblock URL
  \url{http://rspa.royalsocietypublishing.org/content/466/2120/2341}.

\bibitem[Dunkin(1965)]{Dunkin1965}
J.~W. Dunkin.
\newblock Computation of modal solutions in layered, elastic media at high
  frequencies.
\newblock \emph{Bulletin of the Seismological Society of America}, 55\penalty0
  (2):\penalty0 335, 1965.
\newblock URL \url{+ http://dx.doi.org/}.

\bibitem[Farzbod and Leamy(2011)]{Farhad-2011}
F.~Farzbod and M.~J. Leamy.
\newblock Analysis of bloch's method and the propagation technique in periodic
  structures.
\newblock \emph{Journal of Vibration and Acoustics}, 133\penalty0 (3):\penalty0
  031010, 2011.
\newblock \doi{http://dx.doi.org/10.1115/1.4003202}.
\newblock URL \url{http://dx.doi.org/10.1115/1.4003202}.

\bibitem[Graff(1975)]{graff1975wave}
K.F. Graff.
\newblock \emph{Wave Motion in Elastic Solids}.
\newblock Dover Books on Physics Series. Dover Publications, 1975.
\newblock ISBN 9780486667454.
\newblock URL \url{https://books.google.co.il/books?id=5cZFRwLuhdQC}.

\bibitem[Hashin(1983)]{Hashin1983}
Z.~Hashin.
\newblock Analysis of composite materials--a survey.
\newblock \emph{Journal of Applied Mechanics}, 50\penalty0 (3):\penalty0
  481--505, September 1983.
\newblock ISSN 0021-8936.
\newblock \doi{10.1115/1.3167081}.
\newblock URL \url{http://dx.doi.org/10.1115/1.3167081}.

\bibitem[Herzig~Sheinfux et~al.(2014)Herzig~Sheinfux, Kaminer, Plotnik, Bartal,
  and Segev]{2014-Effective-medium-breakdown-Hanan-Segev}
H.~Herzig~Sheinfux, I.~Kaminer, Y.~Plotnik, G.~Bartal, and M.~Segev.
\newblock Subwavelength multilayer dielectrics: Ultrasensitive transmission and
  breakdown of effective-medium theory.
\newblock \emph{Phys. Rev. Lett.}, 113:\penalty0 243901, Dec 2014.
\newblock \doi{10.1103/PhysRevLett.113.243901}.
\newblock URL \url{https://link.aps.org/doi/10.1103/PhysRevLett.113.243901}.

\bibitem[Joseph and Craster(2015)]{joseph2015}
L.M. Joseph and R.V. Craster.
\newblock Reflection from a semi-infinite stack of layers using homogenization.
\newblock \emph{Wave Motion}, 54:\penalty0 145 -- 156, 2015.
\newblock ISSN 0165-2125.
\newblock \doi{https://doi.org/10.1016/j.wavemoti.2014.12.003}.
\newblock URL
  \url{http://www.sciencedirect.com/science/article/pii/S0165212514001747}.

\bibitem[Kittel(2005)]{Kittel}
C.~Kittel.
\newblock \emph{Introduction to solid state physics, 8th ed.}
\newblock John Wiley \& Sons, Inc., 2005.

\bibitem[Korvink and Paul(2006)]{Korvink2006MEMS}
Jan Korvink and Oliver Paul.
\newblock \emph{MEMS: A Practical Guide of Design, Analysis, and Applications}.
\newblock Springer-Verlag Berlin Heidelberg, 2006.
\newblock ISBN 978-3-540-21117-4.
\newblock \doi{http://dx.doi.org/10.1007/978-3-540-33655-6}.

\bibitem[Ma and Sheng(2016)]{2016-acoustic-metamt-broad-horizons}
G.~Ma and P.~Sheng.
\newblock Acoustic metamaterials: From local resonances to broad horizons.
\newblock \emph{Science Advances}, 2\penalty0 (2), 2016.
\newblock \doi{10.1126/sciadv.1501595}.
\newblock URL \url{http://advances.sciencemag.org/content/2/2/e1501595}.

\bibitem[Milton and Willis(2007)]{Milton07}
G.~W. Milton and J.~R. Willis.
\newblock On modifications of newton{\textquoteright}s second law and linear
  continuum elastodynamics.
\newblock \emph{Proceedings of the Royal Society of London A: Mathematical,
  Physical and Engineering Sciences}, 463\penalty0 (2079):\penalty0 855--880,
  2007.
\newblock ISSN 1364-5021.
\newblock \doi{10.1098/rspa.2006.1795}.
\newblock URL
  \url{http://rspa.royalsocietypublishing.org/content/463/2079/855}.

\bibitem[Milton(2002)]{milt02book}
G.W. Milton.
\newblock \emph{The Theory of Composites}, volume~6 of \emph{Cambridge
  Monographs on Applied and Computational Mathematics}.
\newblock Cambridge University Press, New York, 2002.

\bibitem[Misseroni et~al.(2016)Misseroni, Colquitt, Movchan, Movchan, and
  Jones]{Misseroni2016sr}
D.~Misseroni, D.~J. Colquitt, A.~B. Movchan, N.~V. Movchan, and I.~S. Jones.
\newblock Cymatics for the cloaking of flexural vibrations in a structured
  plate.
\newblock \emph{Scientific Reports}, 6:\penalty0 23929 EP --, 04 2016.
\newblock URL \url{http://dx.doi.org/10.1038/srep23929}.

\bibitem[Nemat-Nasser and Hori(1999)]{Nemat-Nasser1999}
S.~Nemat-Nasser and M.~Hori.
\newblock \emph{Micromechanics : overall properties of heterogeneous materials,
  Second edition}.
\newblock Amsterdam: Elsevier, 1999.
\newblock ISBN 0444500847.

\bibitem[Nemat-Nasser and Srivastava(2011)]{NematNasser2011jmps}
S.~Nemat-Nasser and A.~Srivastava.
\newblock Overall dynamic constitutive relations of layered elastic composites.
\newblock \emph{Journal of the Mechanics and Physics of Solids}, 59\penalty0
  (10):\penalty0 1953--1965, 2011.
\newblock ISSN 0022-5096.
\newblock \doi{http://dx.doi.org/10.1016/j.jmps.2011.07.008}.
\newblock URL
  \url{http://www.sciencedirect.com/science/article/pii/S0022509611001475}.

\bibitem[Nemat-Nasser et~al.(2011)Nemat-Nasser, Willis, Srivastava, and
  Amirkhizi]{NN11srivastava}
S.~Nemat-Nasser, J.~R. Willis, A.~Srivastava, and A.~V. Amirkhizi.
\newblock Homogenization of periodic elastic composites and locally resonant
  sonic materials.
\newblock \emph{Phys. Rev. B}, 83:\penalty0 104103, Mar 2011.
\newblock \doi{10.1103/PhysRevB.83.104103}.
\newblock URL \url{https://link.aps.org/doi/10.1103/PhysRevB.83.104103}.

\bibitem[Ostoja-Starzewski(2002)]{Ostoja-Starzewski2002fk}
M.~Ostoja-Starzewski.
\newblock Lattice models in micromechanics.
\newblock \emph{Applied Mechanics Reviews}, 55\penalty0 (1):\penalty0 35--60,
  01 2002.
\newblock URL \url{http://dx.doi.org/10.1115/1.1432990}.

\bibitem[P\'erez-\'Alvarez and Garc\'{\i}a-Moliner(2004)]{Perez2004}
R.~P\'erez-\'Alvarez and F.~Garc\'{\i}a-Moliner.
\newblock \emph{Transfer Matrix, Green Function and related techniques: Tools
  for the study of multilayer heterostructures}.
\newblock Universitat Jaume I, Castell\'{o}n de la Plana, Spain, 2004.
\newblock ISBN 84-8021-474-4.

\bibitem[P{\'e}rez-{\'A}lvarez et~al.(2015)P{\'e}rez-{\'A}lvarez,
  Pernas-Salom{\'o}n, and Velasco]{rene15siam}
R.~P{\'e}rez-{\'A}lvarez, R.~Pernas-Salom{\'o}n, and V.R. Velasco.
\newblock Relations between transfer matrices and numerical stability analysis
  to avoid the $\omega$d problem.
\newblock \emph{SIAM Journal on Applied Mathematics}, 75\penalty0 (4):\penalty0
  1403--1423, 2015.
\newblock \doi{10.1137/140993442}.
\newblock URL \url{http://dx.doi.org/10.1137/140993442}.

\bibitem[Shmuel and Pernas-Salom\'on(2016)]{2016-Shmuel-SMS}
G.~Shmuel and R.~Pernas-Salom\'on.
\newblock Manipulating motions of elastomer films by
  electrostatically-controlled aperiodicity.
\newblock \emph{Smart Materials and Structures}, 25\penalty0 (12):\penalty0
  125012, 2016.
\newblock URL \url{http://stacks.iop.org/0964-1726/25/i=12/a=125012}.

\bibitem[Shuguang et~al.(2015)Shuguang, Tianxin, Xudong, and
  Jialu]{Shuguang2015}
Z.~Shuguang, N.~Tianxin, W.~Xudong, and F.~Jialu.
\newblock Studies of band gaps in flexural vibrations of a locally resonant
  beam with novel multi-oscillator configuration.
\newblock \emph{Journal of Vibration and Control}, 23\penalty0 (10):\penalty0
  1663--1674, 2018/03/08 2015.
\newblock \doi{10.1177/1077546315598032}.
\newblock URL \url{https://doi.org/10.1177/1077546315598032}.

\bibitem[Srivastava(2015)]{2015-Elastic-metamaterials-and-dynamic-homogenization-review}
A.~Srivastava.
\newblock Elastic metamaterials and dynamic homogenization: a review.
\newblock \emph{International Journal of Smart and Nano Materials}, 6\penalty0
  (1):\penalty0 41--60, 2015.
\newblock \doi{10.1080/19475411.2015.1017779}.
\newblock URL \url{https://doi.org/10.1080/19475411.2015.1017779}.

\bibitem[Srivastava and Nemat-Nasser(2014)]{Srivastava2014}
A.~Srivastava and S.~Nemat-Nasser.
\newblock On the limit and applicability of dynamic homogenization.
\newblock \emph{Wave Motion}, 51\penalty0 (7):\penalty0 1045 -- 1054, 2014.
\newblock ISSN 0165-2125.
\newblock \doi{https://doi.org/10.1016/j.wavemoti.2014.04.003}.
\newblock URL
  \url{http://www.sciencedirect.com/science/article/pii/S0165212514000614}.

\bibitem[Sun et~al.(2017)Sun, Zhou, Ichchou, Lain\'e, and
  Zine]{2017-Sun-IJAppMech}
X.~Sun, C.~Zhou, M.~Ichchou, J.-P. Lain\'e, and A.-M. Zine.
\newblock Multi-scale homogenization of transversal waves in periodic composite
  beams.
\newblock \emph{International Journal of Applied Mechanics}, 09\penalty0
  (03):\penalty0 1750039, 2017.
\newblock \doi{10.1142/S1758825117500399}.
\newblock URL
  \url{http://www.worldscientific.com/doi/abs/10.1142/S1758825117500399}.

\bibitem[Tan(2010)]{2010-Tan}
E.~L. Tan.
\newblock Generalized eigenproblem of hybrid matrix for floquet wave
  propagation in one-dimensional phononic crystals with solids and fluids.
\newblock \emph{Ultrasonics}, 50\penalty0 (1):\penalty0 91 -- 98, 2010.
\newblock ISSN 0041-624X.
\newblock \doi{https://doi.org/10.1016/j.ultras.2009.09.007}.
\newblock URL
  \url{http://www.sciencedirect.com/science/article/pii/S0041624X09001085}.

\bibitem[Tan(2006)]{tan06hybrid}
E.L. Tan.
\newblock Hybrid compliance-stiffness matrix method for stable analysis of
  elastic wave propagation in multilayered anisotropic media.
\newblock \emph{The Journal of the Acoustical Society of America}, 119\penalty0
  (1), 2006.
\newblock \doi{http://dx.doi.org/10.1121/1.2139617}.
\newblock URL
  \url{http://scitation.aip.org/content/asa/journal/jasa/119/1/10.1121/1.2139617}.

\bibitem[Torrent et~al.(2014)Torrent, Pennec, and
  Djafari-Rouhani]{torrent2014prb}
D.~Torrent, Y.~Pennec, and B.~Djafari-Rouhani.
\newblock Effective medium theory for elastic metamaterials in thin elastic
  plates.
\newblock \emph{Phys. Rev. B}, 90:\penalty0 104110, Sep 2014.
\newblock \doi{10.1103/PhysRevB.90.104110}.
\newblock URL \url{https://link.aps.org/doi/10.1103/PhysRevB.90.104110}.

\bibitem[Willis(2009)]{Willis2009}
J.R. Willis.
\newblock Exact effective relations for dynamics of a laminated body.
\newblock \emph{Mechanics of Materials}, 41\penalty0 (4):\penalty0 385 -- 393,
  2009.
\newblock ISSN 0167-6636.
\newblock \doi{http://dx.doi.org/10.1016/j.mechmat.2009.01.010}.
\newblock URL
  \url{http://www.sciencedirect.com/science/article/pii/S0167663609000118}.
\newblock The Special Issue in Honor of Graeme W. Milton.

\bibitem[Willis(2013)]{Willis2013}
J.R. Willis.
\newblock Some thoughts on dynamic effective properties--a working document.
\newblock 2013.
\newblock URL \url{https://arxiv.org/abs/1311.3875}.

\bibitem[Xiao et~al.(2013)Xiao, Wen, Yu, and
  Wen]{2013-Xiao-periodic-beams-vibration-absorbers}
Y.~Xiao, J.~Wen, D.~Yu, and X.~Wen.
\newblock Flexural wave propagation in beams with periodically attached
  vibration absorbers: Band-gap behavior and band formation mechanisms.
\newblock \emph{Journal of Sound and Vibration}, 332\penalty0 (4):\penalty0 867
  -- 893, 2013.
\newblock ISSN 0022-460X.
\newblock \doi{https://doi.org/10.1016/j.jsv.2012.09.035}.
\newblock URL
  \url{http://www.sciencedirect.com/science/article/pii/S0022460X12007596}.

\bibitem[Xu et~al.(2016)Xu, Zhou, Wang, Wang, Peng, and Li]{XU2016}
Y.~Xu, X.~Zhou, W.~Wang, L.~Wang, F.~Peng, and B.~Li.
\newblock On natural frequencies of non-uniform beams modulated by finite
  periodic cells.
\newblock \emph{Physics Letters A}, 380\penalty0 (40):\penalty0 3278 -- 3283,
  2016.
\newblock ISSN 0375-9601.
\newblock \doi{https://doi.org/10.1016/j.physleta.2016.07.057}.
\newblock URL
  \url{http://www.sciencedirect.com/science/article/pii/S0375960116304960}.

\bibitem[Yang et~al.({2017})Yang, Kim, Cho, and Park]{Yang2017em}
W.~Yang, B.~Kim, S.~Cho, and J.~Park.
\newblock {Experimental Method to Evaluate Effective Dynamic Properties of a
  Meta-Structure for Flexural Vibrations}.
\newblock \emph{Experimental Mechanics}, {57}\penalty0 ({3}):\penalty0
  {417--425}, {MAR} {2017}.
\newblock ISSN {0014-4851}.
\newblock \doi{{10.1007/s11340-016-0242-2}}.

\bibitem[Yu et~al.(2006)Yu, Liu, Wang, Zhao, and Qiu]{Dianlong2006}
D.~Yu, Y.~Liu, G.~Wang, H.~Zhao, and J.~Qiu.
\newblock Flexural vibration band gaps in timoshenko beams with locally
  resonant structures.
\newblock \emph{Journal of Applied Physics}, 100\penalty0 (12):\penalty0
  124901, 2006.
\newblock \doi{10.1063/1.2400803}.
\newblock URL \url{https://doi.org/10.1063/1.2400803}.

\bibitem[Zareei et~al.(2018)Zareei, Darabi, Leamy, and Alam]{Zareei2018}
A.~Zareei, A.~Darabi, M.~J. Leamy, and Mohammad-Reza Alam.
\newblock Continuous profile flexural grin lens: Focusing and harvesting
  flexural waves.
\newblock \emph{Applied Physics Letters}, 112\penalty0 (2):\penalty0 023901,
  2018.
\newblock \doi{10.1063/1.5008576}.
\newblock URL \url{https://doi.org/10.1063/1.5008576}.

\end{thebibliography}

\end{document}